\documentclass[aps, prd, preprint, a4paper, 12pt, superscriptaddress, nofootinbib]{revtex4-1}
\usepackage{graphicx}
\usepackage{amsmath,amssymb}
\usepackage{comment}
\usepackage{here}
\usepackage{docmute}
\usepackage{simplewick}
\usepackage{layout}
\usepackage[margin=2.4cm]{geometry}
\usepackage{color}

\usepackage[normalem]{ulem} %% memo: if [normalem] is removed, italics in refs.(e.g. et al.) change to underlined romans.
\usepackage{cancel}

\begin{document}

\title{\bf Emergence of the $\rho$ resonance from the HAL QCD potential in lattice QCD}

\author{Yutaro Akahoshi}
\affiliation{Center for Gravitational Physics, Yukawa Institute for Theoretical Physics\\Kyoto University, Kyoto 606-8502, Japan}
\affiliation{RIKEN Nishina Center (RNC), Saitama 351-0198, Japan}

\author{Sinya Aoki}
\affiliation{Center for Gravitational Physics, Yukawa Institute for Theoretical Physics\\Kyoto University, Kyoto 606-8502, Japan}
\affiliation{RIKEN Nishina Center (RNC), Saitama 351-0198, Japan}

\author{Takumi Doi}
\affiliation{RIKEN Nishina Center (RNC), Saitama 351-0198, Japan}
\affiliation{RIKEN Interdisciplinary Theoretical and Mathematical Sciences Program (iTHEMS), Saitama 351-0198, Japan}

\preprint{YITP-21-57} \preprint{RIKEN-QHP-496} \preprint{RIKEN-iTHEMS-Report-21}
%%%%% abstract %%%%%
\begin{abstract}

We investigate the $I=1$ $\pi \pi$ interaction using the HAL QCD method in lattice QCD.
We employ the (2+1)-flavor gauge configurations on $32^3 \times 64$ lattice at the lattice spacing $a \approx 0.0907$ fm and $m_{\pi} \approx 411$ MeV, in which the $\rho$ meson appears as a resonance state.
We find that all-to-all propagators necessary in this calculation
can be obtained with reasonable precision
by a combination of three techniques, the one-end trick, the sequential propagator, and the covariant approximation averaging (CAA).
The non-local $I=1$ $\pi \pi$ potential is determined at the next-to-next-to-leading order (N$^2$LO) of the derivative expansion for the first time,
and the resonance parameters of the $\rho$ meson are extracted.
The obtained $\rho$ meson mass is found to be consistent with the value in the literature,
while the value of the coupling $g_{\rho \pi \pi}$ turns out to be somewhat larger.
The latter observation is most likely attributed to the lack of low-energy information
in our lattice setup with the center-of-mass frame.
Such a limitation may appear in other P-wave resonant systems
and we discuss possible improvement in future.
With this caution in mind,
we positively conclude that we can reasonably extract the N$^2$LO potential and resonance parameters even in the system requiring the all-to-all propagators in the HAL QCD method,
which opens up new possibilities for the study of resonances in lattice QCD.

\end{abstract}
%%%%% abstract end %%%%%
\maketitle

%%%%% introduction %%%%%
\section{Introduction} \label{sect:intro}

% Background: the importance of hadronic resonance studies
Understanding the hadronic resonances from the first-principle lattice QCD simulation is one of the most important subjects in particle and nuclear physics.
At present, the finite-volume method~\cite{Luscher:1990ux,Rummukainen:1995vs,Hansen:2012tf}
and the HAL QCD method~\cite{Ishii:2006ec,Aoki:2009ji,Aoki:2011ep,HALQCD:2012aa} are employed to extract hadron interactions.
The finite-volume method extracts scattering phase shifts through finite-volume energy spectra obtained from temporal correlation functions.
It has been successfully applied for various two-meson interactions and related mesonic resonances~\cite{Briceno:2017max}.
In particular, the $\rho$ resonance has been studied extensively~\cite{Aoki:2011yj,Feng:2010es,Lang:2011mn,Dudek:2012xn,Wilson:2015dqa,Alexandrou:2017mpi,Andersen:2018mau,Werner:2019hxc,Fischer:2020fvl} as a benchmark,
since it is experimentally well-established and is easily investigated by the single-channel approximation of the $I=1$ $\pi \pi$ P-wave scattering.
Recent studies report the results with multiple lattice spacings~\cite{Werner:2019hxc}, and those with pion masses including physical masses~\cite{Fischer:2020fvl}.

% HAL (modified)
The HAL QCD method directly constructs inter-hadron potentials from spatial and temporal correlation functions calculated in lattice QCD.
In this method, the potentials can be extracted even without ground state saturations for correlation functions~\cite{HALQCD:2012aa}.
Scattering parameters are then obtained by solving the Schr\"odinger equation in infinite-volume without any model-dependent ansatz.
These features make this method particularly useful to study baryonic systems~\cite{Iritani:2018vfn} and coupled channel systems~\cite{Aoki:2011gt}.
The HAL QCD method has been successfully applied to many hadronic systems
(see Ref.~\cite{Aoki:2020bew} and references therein for the recent status),
including the detailed coupled channel studies for the tetraquark candidate $Z_c(3900)$~\cite{Ikeda:2016zwx,Ikeda:2017mee}
and $H$-dibaryon~\cite{Sasaki:2019qnh}.

There exists, however, a practical challenge in the HAL QCD method
when expanding the scope to many other resonances,
since the expensive computations of all-to-all quark propagators are necessary in most cases.
To overcome this difficulty,
we have previously explored two different all-to-all techniques, the LapH method\cite{Peardon:2009gh} and the hybrid method\cite{Foley:2005ac}.
It turned out~\cite{Kawai:2017goq,Kawai:2018hem}
that the LapH smearing on sink operators with a small number of LapH vectors
enhances non-locality of the HAL QCD potential
and thus the systematic errors associated with the truncation of the derivative expansion.
Increasing the number of LapH vectors to minimize such non-locality
is practically impossible for larger volumes.
On the other hand, the hybrid method with local sink operators is free from
the enhancement of the non-locality and is more suitable for the HAL QCD method.
A series of studies with the hybrid method~\cite{Akahoshi:2019klc,Akahoshi:2020ojo}, however,
has revealed that it requires too much numerical cost for a reduction of stochastic noises to perform large-scale simulations for hadronic resonances.

% about this paper
In this paper, we develop a new strategy to handle all-to-all propagators, where we combine three techniques in lattice QCD, the one-end trick\cite{McNeile:2006bz}, the sequential propagator calculation\cite{Martinelli:1988rr} and the covariant approximation averaging (CAA)\cite{Shintani:2014vja}.
We calculate the HAL QCD potential of the $I=1$ $\pi \pi$ scattering on gauge configurations at $m_{\pi} \approx 411$ MeV, where the $\rho$ meson is known to appear as a resonance with $m_{\rho} \approx 892$ MeV\cite{Aoki:2011yj}.
Numerical accuracy in our new strategy allows us to determine
the non-local $I=1$ $\pi \pi$ potential at the next-to-next-to-leading order (N$^2$LO) in the derivative expansion for the first time.
Accordingly, resonance parameters are extracted also by the N$^2$LO analysis.
A resonance mass of the $\rho$ meson is found to be consistent with previous studies,
while a somewhat larger value of the $\rho\pi\pi$ coupling is obtained.
The latter discrepancy
is most likely attributed to the lack of low-energy information in our lattice setup with
the center-of-mass frame,
indicating that calculations with the laboratory frame\cite{Aoki:2020cwk} in addition to the center-of-mass frame are desirable to study generic P-wave systems in future.

% Introducing the contents of this paper
This paper is organized as follows. In Sect.~\ref{sect:halqcdmethod}, we briefly introduce the HAL QCD method and explain correlation functions relevant to our calculation.
Sect.~\ref{sect:simulationdetails} summarizes simulation details.
In Sect.~\ref{sect:result}, we first present results from the leading order (LO) analysis for two different source operators.
Then, as our main results, we give the N$^2$LO order potential and resonance parameters, which are compared with the previous results in the finite-volume method.
Sect.~\ref{sect:summary} is devoted to a summary of this study.
In Appendix~\ref{appex:oneend}, we explain the one-end trick, which is a clever way to treat a certain combination of all-to-all propagators.
Details of calculations of quark contraction diagrams are given in Appendix~\ref{appex:diagrams}, while
possible effects of smeared quarks for sink operators are investigated in Appendix~\ref{appex:smrdsink}.
Some details on the N$^2$LO analysis, namely the assumption in the potential fit and behavior of our N$^2$LO potential in terms of energy-dependent local manner, are given in Appendix~\ref{appex:fitassumption} and \ref{appex:enedeplocalpot}.

%%%%% method %%%%%
\section{The HAL QCD method} \label{sect:halqcdmethod}

A fundamental quantity in the HAL QCD method is the Nambu-Bethe-Salpeter (NBS) wave function, which is defined as
\begin{equation}
  \psi_{W}({\bf r}) = \langle 0| (\pi \pi)_{I=1,I_z=0}({\bf r},0) |\pi \pi;I=1,I_z=0,{\bf k} \rangle,
\end{equation}
where $|\pi \pi;I=1,I_z=0,{\bf k} \rangle$ is an asymptotic state for an elastic $I=1$ $\pi \pi$ system in the center-of-mass frame with a relative momentum ${\bf k}$, a total energy $W = 2 \sqrt{m_{\pi}^2 + k^2}$ and $ k = \vert {\bf k} \vert$.
The operator $(\pi \pi)_{I=1,I_z=0}({\bf r},t)$ is a two-pion operator projected to the $I=1, I_z=0$ channel given by
\begin{eqnarray} \label{eq:sinkop}
  (\pi \pi)_{I=1,I_z=0}({\bf r},t) &=& \frac{1}{\sqrt{2}} \{ \pi_s^{+}({\bf r+x},t) \pi_s^{-}({\bf x},t) - \pi_s^{-}({\bf r+x},t) \pi_s^{+}({\bf x},t) \}, \\
  \pi_s^{+}({\bf x},t) &=& \bar d_s ({\bf x},t) \gamma_5 u_s ({\bf x},t), \quad
  \pi_s^{-}({\bf x},t) = \bar u_s ({\bf x},t) \gamma_5 d_s ({\bf x},t),
\end{eqnarray}
where $u_s, d_s$ are smeared up and down quark fields. A detail of the quark smearing is given in Sect.~\ref{sect:simulationdetails}.
A radial part of the $l$-th partial component in the NBS wave function behaves at large $r = \vert{\bf r}\vert$ as~\cite{Aoki:2009ji,Aoki:2013cra}
\begin{equation}
  \psi^{l}_{W}(r) \approx A_{l}({\bf k}) e^{i\delta_l(k)} \frac{\sin(kr-l \pi/2+\delta_l(k))}{kr},
\end{equation}
where $A_{l}({\bf k})$ is an overall factor and $\delta_l(k)$ is a scattering phase shift,
which is equal to a phase of the S-matrix implied by its unitarity.
By using this property, we can construct an energy-independent non-local potential $U({\bf r},{\bf r'})$ as
\begin{equation}
  \frac{1}{2 \mu}(\nabla^2 + k^2) \psi_{W}({\bf r}) = \int d^3 {\bf r'}\, U({\bf r},{\bf r'})\psi_{W}({\bf r'}),
\end{equation}
with $\mu = m_{\pi}/2$ a reduced mass of two pions.
In general, the HAL QCD potential depends on a choice of hadron operators in the NBS wave function (the sink operator $(\pi \pi)_{I=1,I_z=0}({\bf r},t)$ in our case), and it is referred to as the scheme dependence of the potential~\cite{Aoki:2012tk,Kawai:2017goq}.
Physical observables extracted from potentials in different schemes, of course, agree with each other
by construction.
Therefore, we can utilize this scheme dependence to reduce
statistical and/or systematic uncertainties
of observables.
As discussed in Appendix~\ref{appex:smrdsink},
a comparison of different schemes shows that
the $I=1$ $\pi \pi$ potential has much smoother ${\bf r}$ dependences in  an ``equal-time smeared-sink scheme'' (Eq.(\ref{eq:sinkop})), where
sink quark fields are slightly smeared and two pion operators are put on the same time slice.
We employ this scheme for the whole analysis in this study.

To extract the potential in lattice QCD simulations, we begin with a normalized correlation function defined as
\begin{equation}
  R({\bf r},t) \equiv \frac{F_{\pi \pi}({\bf r},t)}{F_{\pi}(t)^2},
\end{equation}
where $F_{\pi}$ and $F_{\pi \pi}$ are a single-pion and a two-pion correlation function, respectively,
\begin{eqnarray}
  F_{\pi}(t) &=& \sum_{{\bf x,y},t_0}\langle \pi^{-}({\bf x},t+t_0) \pi^{+}({\bf y},t_0) \rangle, \\
  F_{\pi \pi}({\bf r},t) &=& \sum_{t_0} \langle (\pi \pi)_{I=1,I_z=0}({\bf r},t+t_0) \overline{ {\mathcal J} }^{T_1^-}_{I=1,I_z=0}(t_0) \rangle.
\end{eqnarray}
Here $\overline{ {\mathcal J} }^{T_1^-}_{I=1,I_z=0}(t_0)$ is a source operator which creates
$\pi \pi$ scattering states with $(I,I_z) =(1,0)$  in an irreducible representation $T_1^-$ of the cubic group.
Thus $R({\bf r},t)$ is related to the NBS wave function as
\begin{equation}
  R({\bf r},t) = \sum_n {B_n} \psi_{W_n}({\bf r}) e^{-(W_n - 2m_{\pi}) t} + \cdots ,
\end{equation}
where $W_n$ and $B_n$ are energy and overlap factor of the $n$-th elastic state, and
the ellipses indicate inelastic contributions. Using the energy independence of the potential, we can show that\cite{HALQCD:2012aa}
\begin{equation} \label{eq:timedepHAL}
  \left[ \frac{\nabla^2}{2 \mu} -\frac{\partial}{\partial t} + \frac{1}{8 \mu} \frac{\partial^2}{\partial t^2} \right] R({\bf r},t) = \int d^3{\bf r'} U({\bf r},{\bf r'}) R({\bf r'},t),
\end{equation}
at a sufficiently large $t$ where inelastic contributions in $R({\bf r},t)$ becomes negligible.
In actual calculations, we {introduce a} derivative expansion to treat the non-local potential as
\begin{equation} \label{eq:derivexp}
  U ({\bf r},{\bf r'}) = (V_0 (r) + V_2 (r) \nabla^2 + {\mathcal O}(\nabla^4)) \delta({\bf r-r'}),
\end{equation}
and the effective LO potential is given by
\begin{equation}
  V^{\rm LO}(r) = \frac{ \sum_{g\in O_h} R^{\dag}(g{\bf r},t) \left[ \dfrac{\nabla^2}{2 \mu} -\dfrac{\partial}{\partial t} + \dfrac{1}{8 \mu} \dfrac{\partial^2}{\partial t^2} \right] R(g{\bf r},t)}{\sum_{g\in O_h} R^{\dag}(g{\bf r},t) R(g{\bf r},t)},
\end{equation}
where invariance of the potential under the cubic rotation group $O_h$ is utilized to improve signals~\cite{Murano:2013xxa}.
In this study, we further determine the effective N$^2$LO potential in order to extract resonance parameters more accurately.
The effective N$^2$LO potential $U^{\rm N^2LO}({\bf r},{\bf r'}) = \left( V_0^{\rm N^2LO} + V_2^{\rm N^2LO}  \nabla^2 \right) \delta({\bf r}-{\bf r'}) $ is determined by {solving} the following linear equations~\cite{Iritani:2018zbt}:
\begin{equation} \label{eq:NLOlineqs}
  \left(
  \begin{array}{cc}
    1 & \nabla^2 R_A ({\bf r},t)/R_A ({\bf r},t) \\
    1 & \nabla^2 R_B ({\bf r},t)/R_B ({\bf r},t)
  \end{array}
  \right)
  \left(
  \begin{array}{c}
    V^{\rm N^2LO}_0(r) \\
    V^{\rm N^2LO}_2(r)
  \end{array}
  \right) =
  \left(
  \begin{array}{c}
    V^{\rm LO}_A(r) \\
    V^{\rm LO}_B(r)
  \end{array}
  \right),
\end{equation}
where $R_i\ (i=A,B)$ are the normalized correlation functions with different source operators $\overline{ {\mathcal J} }_i(t_0)\ (i=A,B)$, and $V^{\rm LO}_i(r)\ (i=A,B)$ are the effective LO potentials obtained by $R_i\ (i=A,B)$.
% implicit dependence of input energy
Note that coefficients $V_0(r), V_2(r)$ in the full derivative expansion (Eq.(\ref{eq:derivexp})) are independent of source operators, while effective N$^2$LO {coefficients} $V_0^{\rm N^2LO}(r), V_2^{\rm N^2LO}(r)$ depend on a choice of source operators.
In other words, effective potentials implicitly depend on discrete energy levels included in their determination due to the truncation of the derivative expansion~\cite{Aoki:2020bew}.
Therefore,
systematic errors in the derivative expansion for physical observables depend on
the magnitude of non-locality in the true potential
as well as on the difference between the energy region relevant for physical observables
and that employed to determine the effective potentials~\cite{Aoki:2020bew}.

For the source operators, we choose $\rho$-type $\overline{ {\mathcal J} }^{T_1^-}_{\rho, I=1,I_z=0}(t_0)$ and $\pi\pi$-type $\overline{ {\mathcal J} }^{T_1^-}_{\pi\pi, I=1,I_z=0}(t_0)$ in this study, defined by
\begin{eqnarray}
  \overline{ {\mathcal J} }^{T_1^-}_{\rho, I=1,I_z=0}(t_0) &=& \overline{\rho}^{0}_3 (t_0),\\
  \overline{ {\mathcal J} }^{T_1^-}_{\pi\pi, I=1,I_z=0}(t_0) &=&
  %{\mathcal P}^{T_1^-} \overline{(\pi \pi)}_{I=1,I_z=0}({\bf p}_3,t_0) =
  \overline{(\pi \pi)}_{I=1,I_z=0}({\bf p}_3,t_0) ,
\end{eqnarray}
%where ${\mathcal P}^{T_1^-}$ is a projection operator onto the irreducible representation $T_1^-$ of the cubic group and
where ${\bf p}_3 = (0,0,2\pi/L)$. $\overline{(\pi\pi)}_{I=1,I_z=0}({\bf p},t)$ and $\overline{\rho}^{0}_3$ are given as
\begin{eqnarray}
  \overline{\rho}^{0}_3 (t) &=& \sum_{{\bf x}} \frac{1}{\sqrt{2}} \left(  \bar u({\bf x},t) \gamma_{3} u({\bf x},t) - \bar d({\bf x},t) \gamma_{3} d({\bf x},t) \right)\\
  \overline{(\pi\pi)}_{I=1,I_z=0}({\bf p},t) &=& \frac{1}{\sqrt{2}} \sum_{{\bf y_1,y_2}} e^{-i{\bf p \cdot y_1}} e^{i{\bf p \cdot y_2}} \left( \pi^{-}({\bf y_1},t)\pi^{+}({\bf y_2},t) - \pi^{+}({\bf y_1},t)\pi^{-}({\bf y_2},t) \right),
\end{eqnarray}
where we use local quark fields for source operators.

Calculations of correlation functions with momentum projected sources generally need all-to-all propagators, which requires too much numerical cost to calculate exactly.
Therefore, we evaluate all-to-all propagators by the
combination of the one-end trick, the sequential propagator, and the CAA.
We give a brief introduction of the one-end trick in Appendix~\ref{appex:oneend}, and details of diagram calculations are
presented in Appendix~\ref{appex:diagrams}.

%%%%% simulation details %%%%%
\section{Simulation details} \label{sect:simulationdetails}

We employ (2+1)-flavor full QCD configurations generated by the PACS-CS Collaborations~\cite{Aoki:2008sm} on a $32^3 \times 64$ lattice with the Iwasaki gauge action\cite{Iwasaki:1985we} at $\beta=1.90$
and a non-perturbatively improved Wilson-clover action\cite{Sheikholeslami:1985ij} at $c_{SW} = 1.715$ and hopping parameters $(\kappa_{ud},\kappa_s) = (0.13754,0.13640)$.
These parameters correspond to a lattice spacing $a = 0.0907$ fm,
and a pion mass $m_{\pi} \approx 411$ MeV, where the $\rho$ meson appears as a resonance with $m_{\rho} \approx 892$ MeV~\cite{Aoki:2011yj}.
The calculations are performed in the center-of-mass frame with
the periodic boundary condition for all spacetime directions.
In this report, dimensionful quantities without the corresponding unit are written in lattice unit unless otherwise stated.

\begin{table}[tbp]
  \caption{Numerical setup for the calculation.}
  \vspace{2mm}
  \centering
  \begin{tabular}{c|ccc}
    Source type & Scheme & $N_{\rm conf}$ (\#. of time slice ave.) & Stat. error   \\ \hline \hline
    $\pi \pi$-type & equal-time, smeared-sink & 100 (64) & jackknife with bin--size 5 \\
    $\rho$-type & equal-time, smeared-sink & 200 (64) & jackknife with bin--size 10 \\
  \end{tabular}
  \label{tab:setups_general}
\end{table}
\begin{table}[tbp]
  \caption{Setups for the one-end trick and the CAA in this study. $N_{\rm eig}$ is the number of low eigenmodes. Color and spinor dilutions are always used.}
  \vspace{2mm}
  \centering
    \begin{tabular}{c|cc|cc}
      Source type  & \multicolumn{2}{|c|}{One-end trick} & \multicolumn{2}{|c}{CAA} \\
      & Noise vector & Space dilution & \ $N_{\rm eig}$ & \# of averaged points \\ \hline \hline
      $\pi\pi$-type & $Z_4$ noise & $s2$ (even-odd) & \ 300 & 64 \\
      $\rho$-type & $Z_4$ noise & $s4$ & \ 300 & 64
    \end{tabular}
    \label{tab:setups_techniques}
\end{table}

Table \ref{tab:setups_general} and \ref{tab:setups_techniques} show general setups and  parameters of the one-end trick and the CAA, respectively.
% details of smearing
We employ smeared quark operators $q_s({\bf x},t) = \sum_{\bf y}f({\bf x-y}) q({\bf y},t)$ at the sink with the Coulomb gauge fixing, in order to improve signals of potentials {at short distance}. A smearing function $f$ is given by
\begin{equation}
  f \left( {\bf x}  \right) = \begin{cases}
    A e^{-B |{\bf x }|} & ( \ 0 < |{\bf x }| < R \ ) \\
    1 & ( \ |{\bf x }| = 0 \ ) \\
    0 & ( \ |{\bf x }| \geq R \ ),
  \end{cases}
\end{equation}
with $A = 1.0,\ B = 1.0,\ R = 3.5$.
As discussed in Appendix~\ref{appex:smrdsink},
these parameters make potential smoother without worsening the convergence of the derivative expansion.
% noise vectors
For the one-end trick, we generate a single $Z_4$ noise vector for each insertion.
To suppress the corresponding stochastic noises, we employ a dilution technique~\cite{Foley:2005ac} in color, spinor and space indices.
Color and spinor indices are fully diluted, and for the space dilution,
we take $s2$ (even-odd) dilution and $s4$ dilution~\cite{Akahoshi:2020ojo} in the $\pi\pi$-type source and the $\rho$-type source, respectively.
% comment on the CAA
In the CAA, we exactly estimate a low-mode part with 300 eigenmodes, and a high-mode part is estimated by an average over loosely solved solutions on 64 different spatial points ${\bf x} = (x_0 + 8l, y_0 + 8m, z_0 + 8n)\ {\rm mod}\ 32$, with $l,n,m \in \{0,1,2,3\}$. Finer and looser solutions are
obtained with $1.0 \times 10^{-24}$ and $9.0 \times 10^{-9}$ for the squared residue, respectively.
We randomly choose the reference point ${\bf x}_0 = (x_0,y_0,z_0)$ for each configuration.

% 2pt correlation function
\begin{figure}
  \includegraphics[width=80mm,clip]{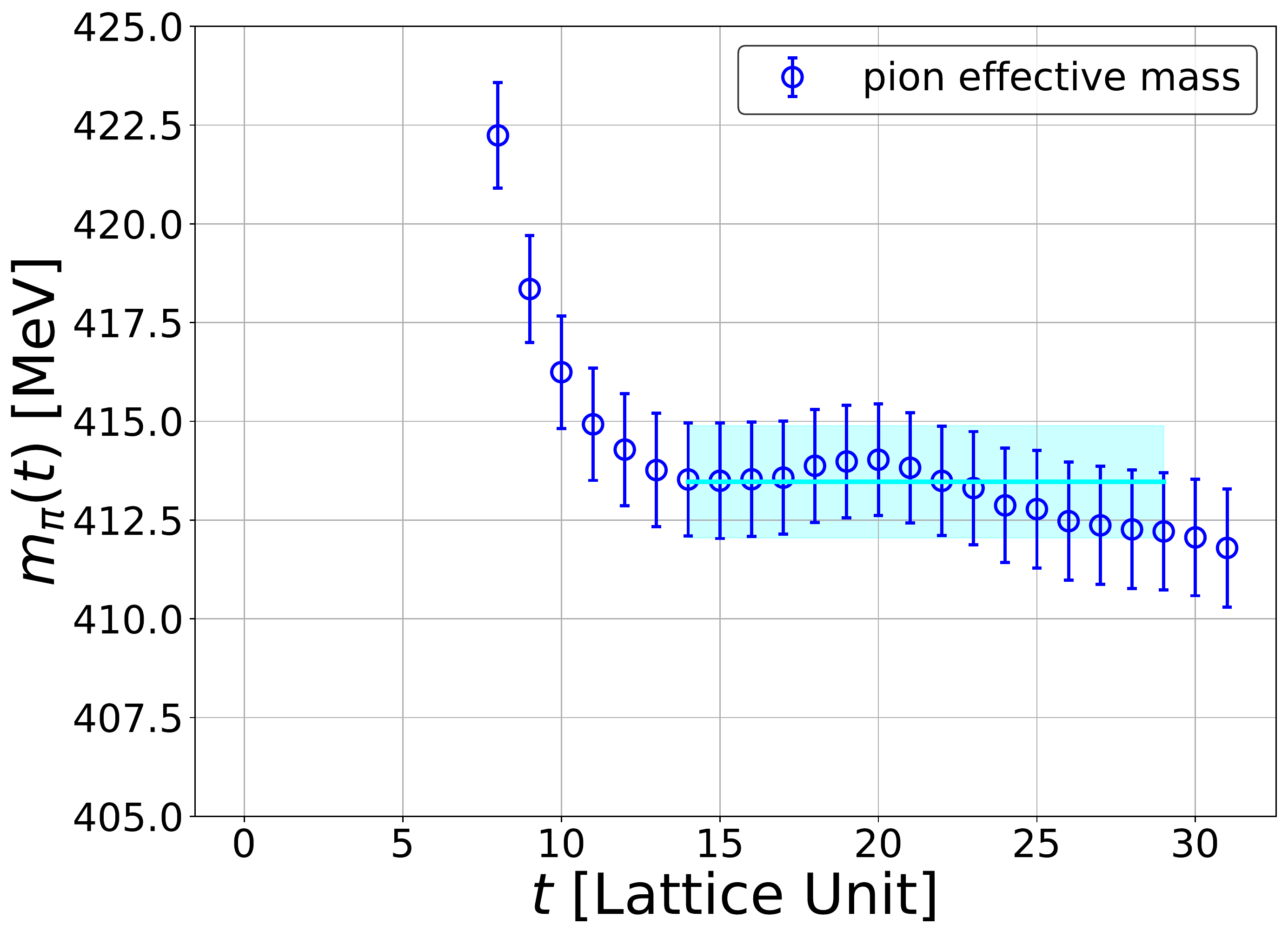}
  \caption{Effective mass of a pion (blue circles) and the fit result by a cosh function at $t = [t_{\rm min},t_{\rm max}] = [14,29]$ (cyan solid line with bands).}
  \label{fig:effmass_pi}
\end{figure}
Figure~\ref{fig:effmass_pi} (left) show an effective mass of a pion obtained by an average over 200 configurations ($\times 64$ time slice average).
A fit to the pion propagator $F_{\pi}(t)$ at $t = [t_{\rm min},t_{\rm max}] = [14,29]$ with a cosh function gives $m_{\pi} = 413.5(1.4)$ MeV.
We also check a $t_{\rm min}$ dependence of the effective mass, and the dependence is negligible compared with statistical errors
as far as $t_{\rm min} \geq 13$.
Therefore we confirm that a ground state saturation in $F_{\pi}(t)$ is achieved at $t=13$.
A possible leading inelastic contribution for two pions in this setup comes from a P-wave $K \overline K$ state with energy $W_{K \overline K} = 2 \sqrt{m_{K}^2 + (2 \pi/L)^2} \approx 1530$ MeV in non-interacting case,
while the two-pion ground state energy is reported as $E_0 = 914(11)$ MeV in Ref.~\cite{Aoki:2011yj}.
We therefore expect inelastic contributions in $F_{\pi \pi}({\bf r},t)$ are suppressed
at $t \approx 1/ [W_{K \overline K} - E_0] \approx 3.5$.
These considerations suggest that inelastic contributions in $R({\bf r},t)$ become negligible at $t \geq 13$, so that potentials can be reliably extracted at $t\geq 13$.
Hereafter, we show results at $t = 14$ and $18$ for $\rho$-type source and $\pi\pi$-type source, respectively.

% parameters for Misner's method
In lattice QCD, the rotational symmetry is broken to the cubic symmetry,
and there exist higher partial wave components in the irreducible representation of the cubic group
($l=3,5,\cdots$ partial waves in this study).
This leads to systematic uncertainties in the HAL QCD potential,
which exhibit as multi-valued structures of potentials as a function of $r$.
We address this issue by performing the approximated partial wave decomposition recently introduced
to lattice QCD~\cite{Miyamoto:2019jjc}.
In practice, we remove the dominant contaminations, the $l=3$ partial wave component,
when we evaluate the potential at $r = [2, 14.8]$.
Tunable parameters of the decomposition~\cite{Miyamoto:2019jjc},
a number of radial bases {$n_{\rm max}$}, a number of partial waves considered {$l_{\rm max}$}
and a width of the shell {$\Delta$}, are taken as $(n_{\rm max}, l_{\rm max}, \Delta) = (4, 5, 1.2)$
at $ 2\le r \le 10$ or $(4,5,1.5)$ at $r > 10$,
where we use larger $\Delta$ at larger $r$ to avoid artificial oscillation of decomposed data
due to too small $\Delta$.

%%%%% result %%%%%
\section{Result} \label{sect:result}
%% LO analysis %%
\subsection{Effective leading-order potentials}
% LO potentials (overview, time dep)
\begin{figure}[htbp]
  \hspace{-10mm}
  \begin{tabular}{cc}
  \begin{minipage}{0.5\hsize}
    \includegraphics[width=80mm,clip]{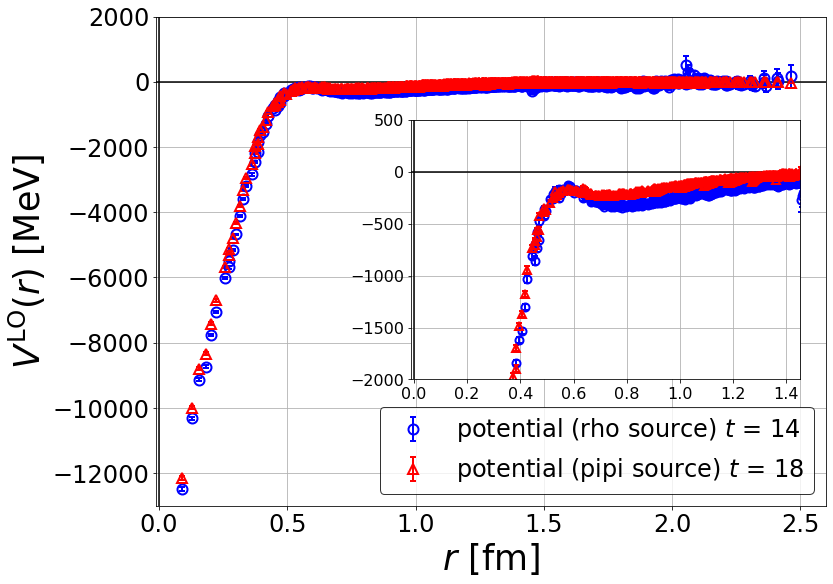}
  \end{minipage} &
  \begin{minipage}{0.5\hsize}
    \includegraphics[width=80mm,clip]{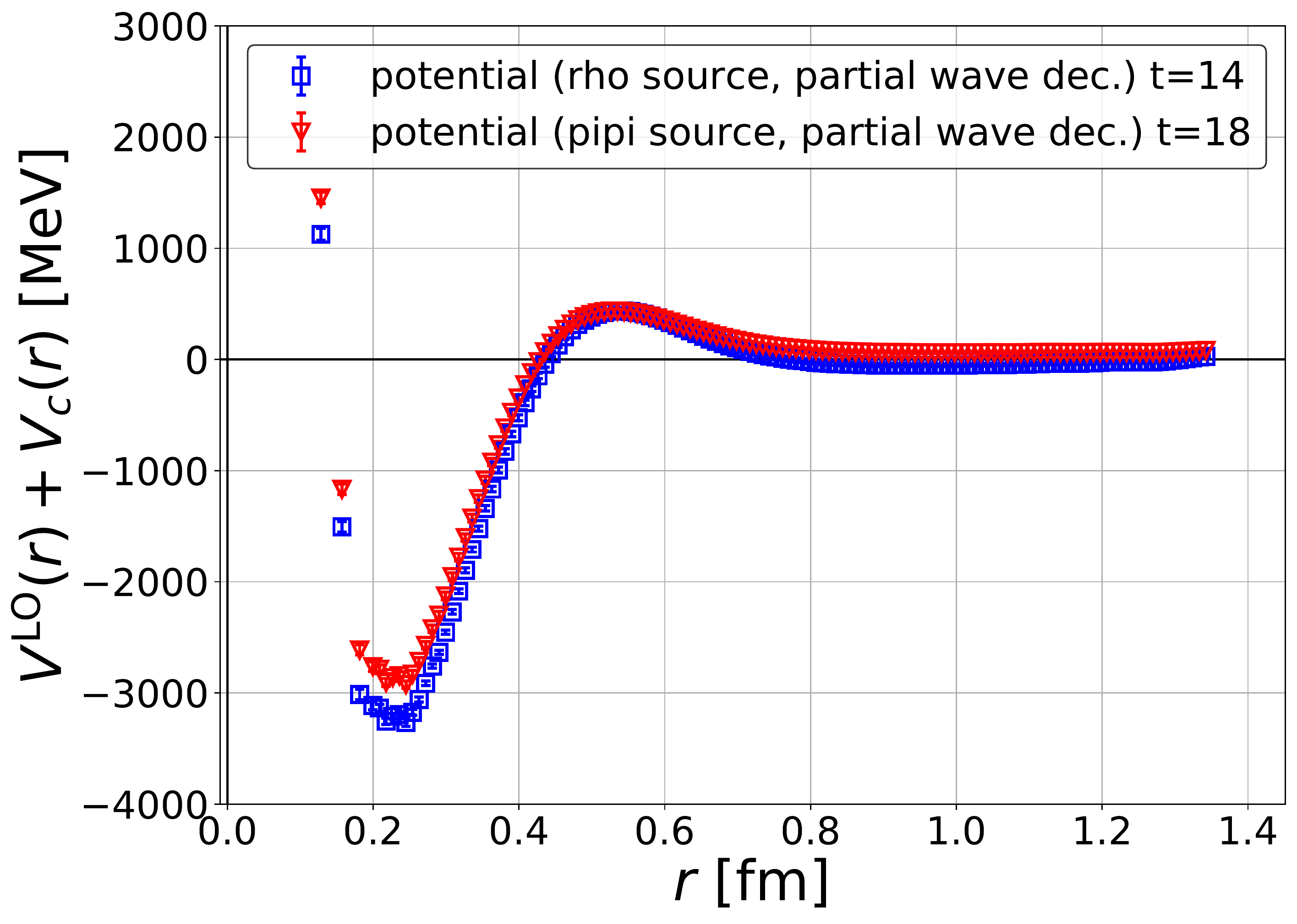}
  \end{minipage}
\end{tabular}
\caption{(Left) Effective LO potentials. Blue and red points show the results from the $\rho$-type source and the $\pi\pi$-type source, respectively. Inset shows an enlarged view of potentials. {(Right) Improved potentials obtained by the partial wave decomposition with the P-wave centrifugal term, $V_c(r) = \frac{1}{2 \mu} \frac{1 \cdot 2}{r^2}$.}}
\label{fig:LOpotentials_t14}
\end{figure}
Figure~\ref{fig:LOpotentials_t14} (Left) show the results for effective LO potentials
without the partial wave decomposition.
We observed that the potentials are attractive at all distances.
Fig.~\ref{fig:LOpotentials_t14} (Right) represents
potentials after the partial wave decomposition with the P-wave centrifugal term added, which become much smoother as multi-valued structures are eliminated.
The potentials with the centrifugal term reveal characteristic features for an existence of a resonance state such as an attractive pocket at short distances and a potential barrier around $r=0.5$ fm.
{We also notice that}  potentials obtained from different source operators {are different from} each other, which fact {suggests a presence of} non-negligible higher-order contributions in the derivative expansion.
\begin{figure}[htbp]
  \hspace{-10mm}
  \begin{tabular}{cc}
  \begin{minipage}{0.5\hsize}
    \includegraphics[width=80mm,clip]{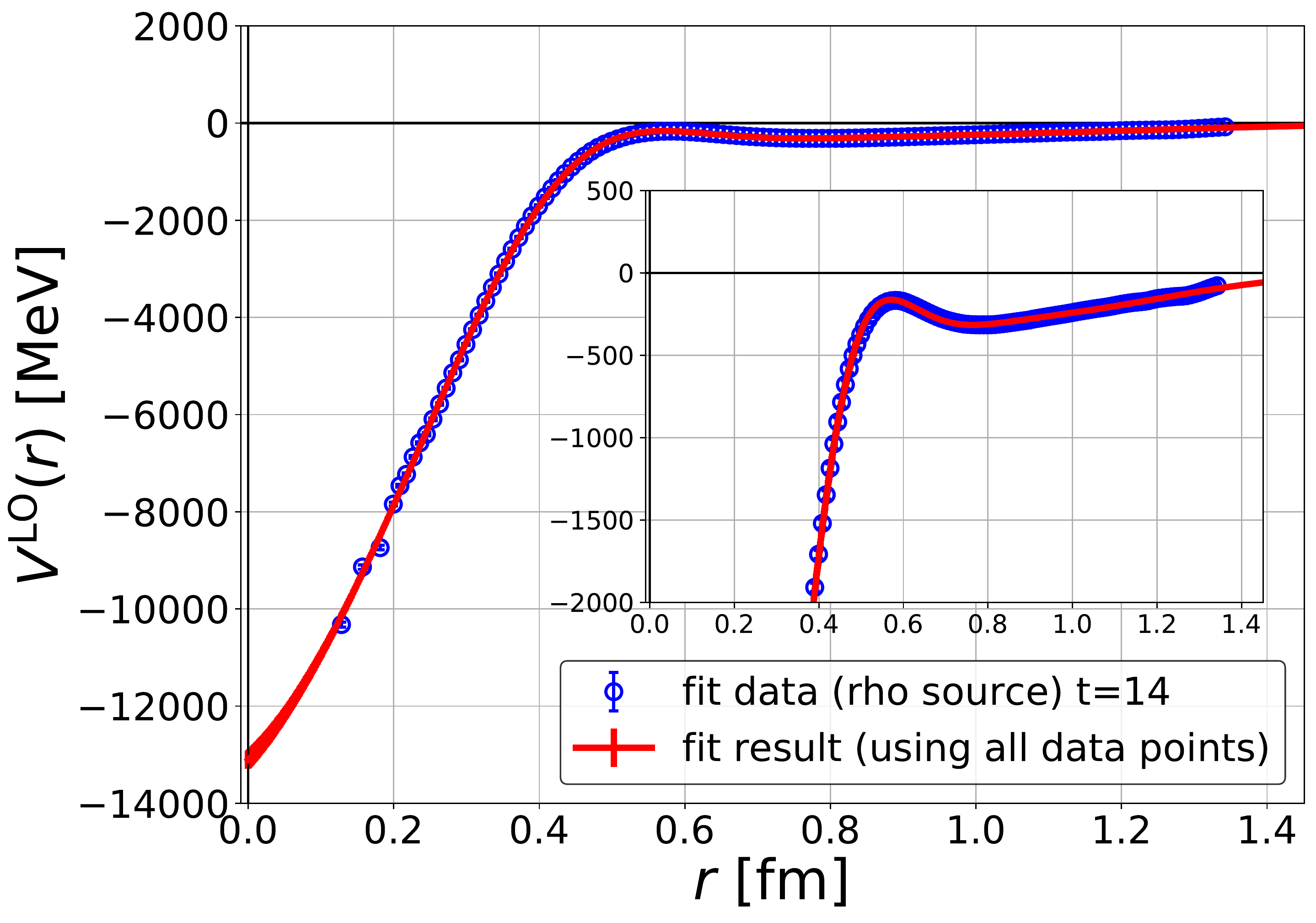}
  \end{minipage} &
  \begin{minipage}{0.5\hsize}
    \includegraphics[width=80mm,clip]{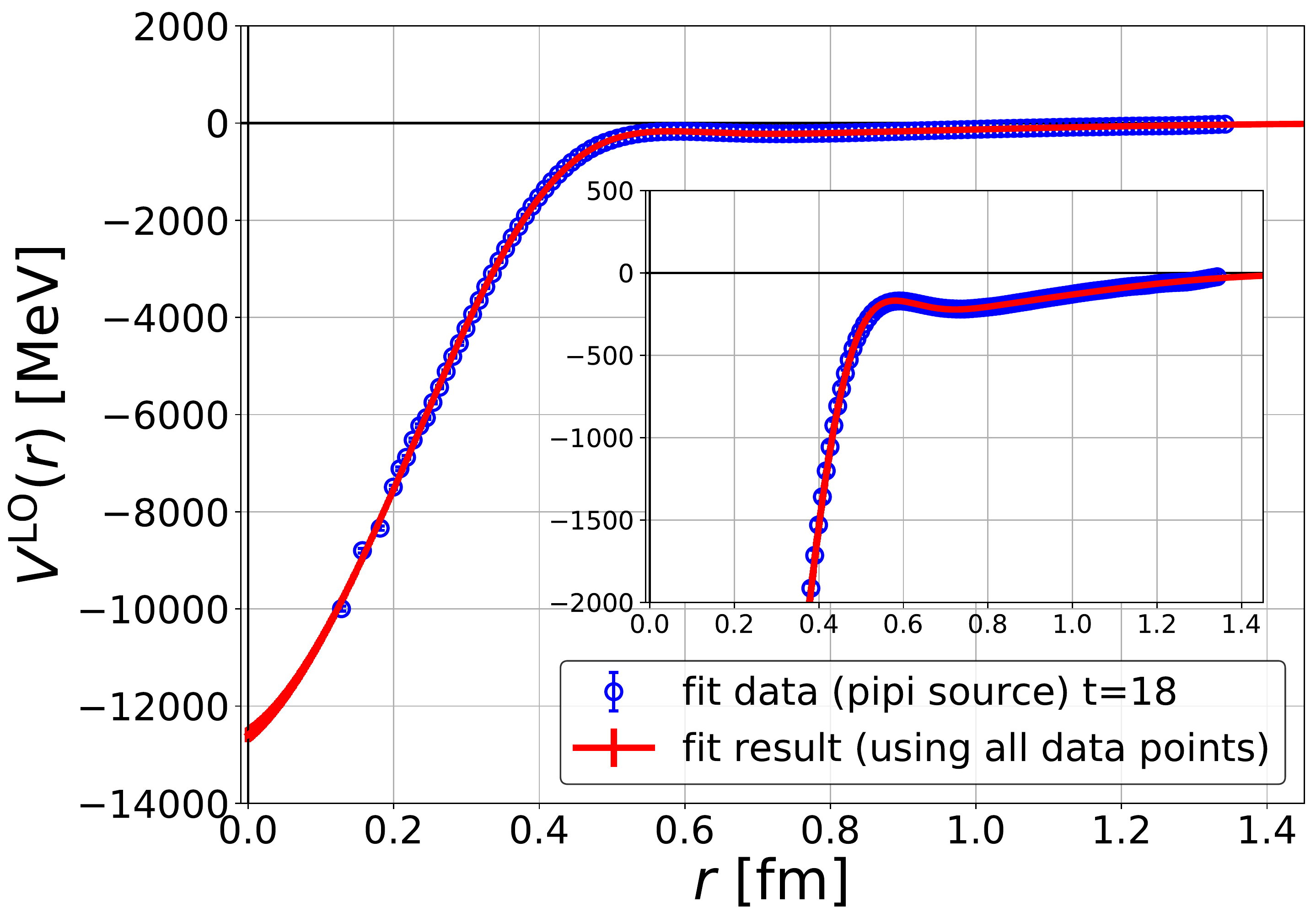}
  \end{minipage}
\end{tabular}
\caption{(Left) Fit result with the $\rho$-type source. Inset shows an enlarged view of them. (Right) The same plot with the $\pi\pi$-type source. Both results are obtained with all allowed data points. }
\label{fig:LOpotentials_fitresult}
\end{figure}
\begin{figure}[htbp]
  \hspace{-10mm}
  \begin{tabular}{cc}
  \begin{minipage}{0.5\hsize}
    \includegraphics[width=80mm,clip]{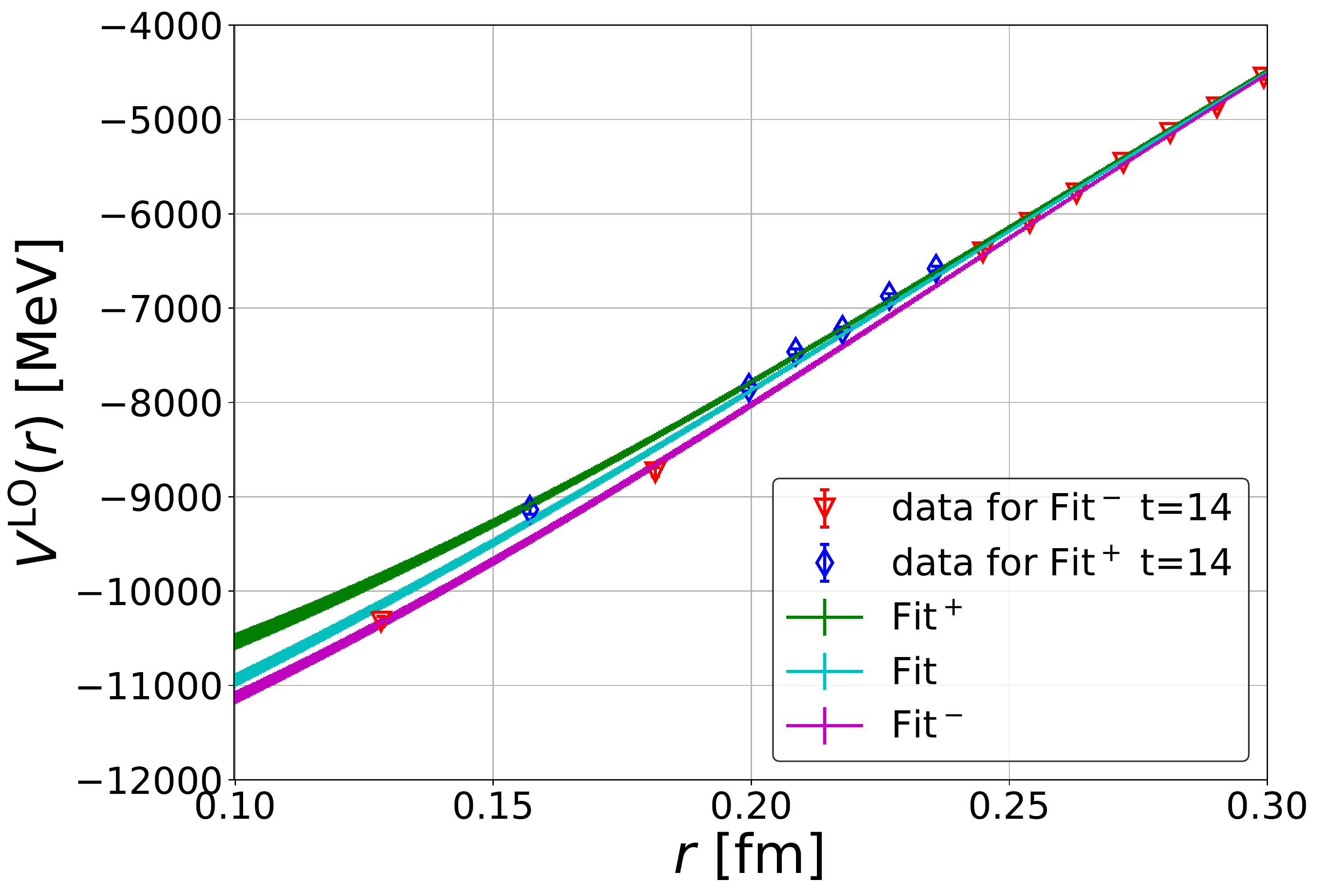}
  \end{minipage} &
  \begin{minipage}{0.5\hsize}
    \includegraphics[width=80mm,clip]{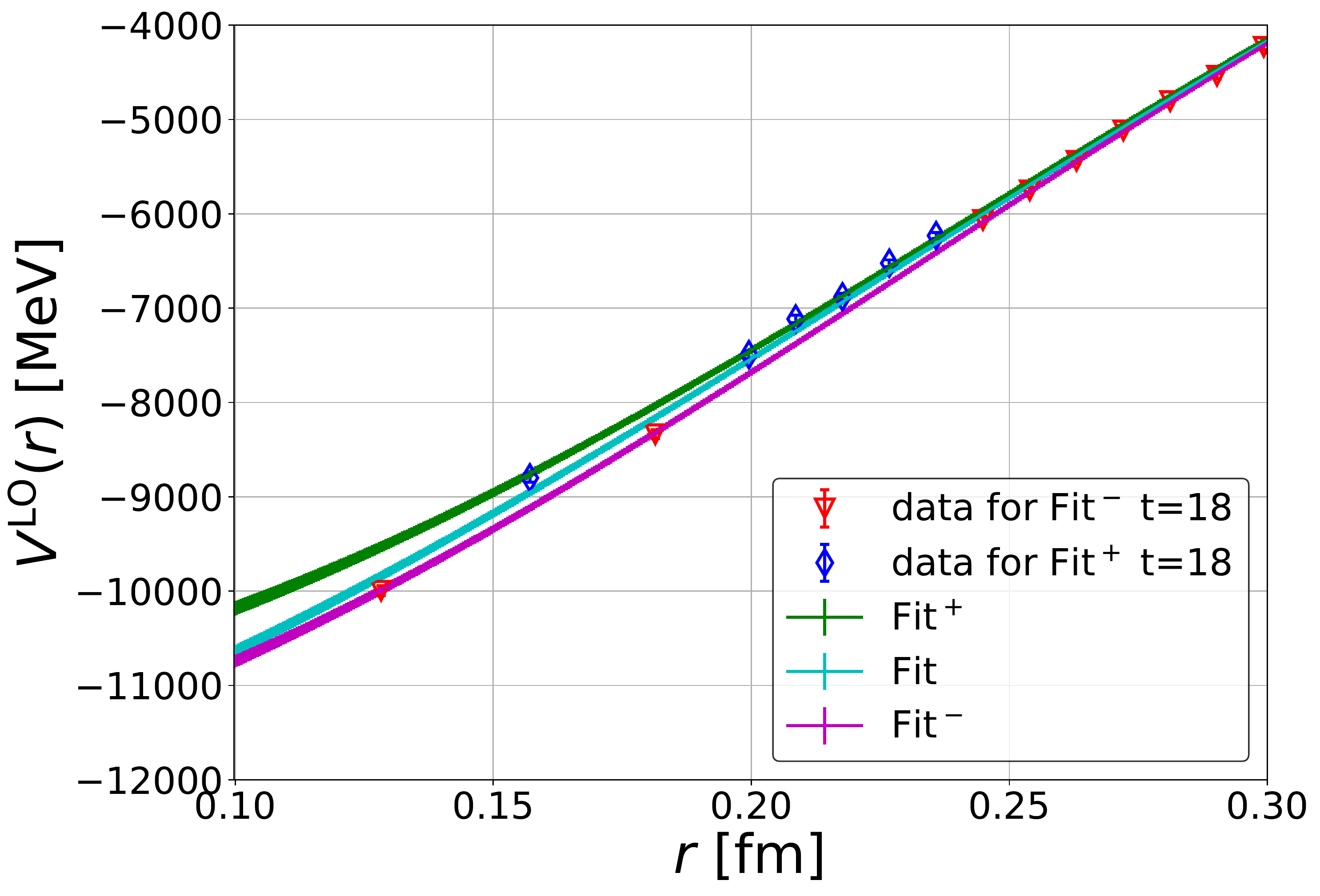}
  \end{minipage}
\end{tabular}
\caption{Systematic uncertainty in the fit of the potential at short distances. (Left) Three fit results {with} the $\rho$-type source. Red and blue points show data used in Fit$^-$ and Fit$^+$, respectively, and magenta and green lines are results of Fit$^-$ and Fit$^+$. We also show the fit result with all allowed data (Fit) by a cyan line for a comparison. (Right) The same plot with the $\pi\pi$-type source.
  }
\label{fig:LOpotentials_fitsys}
\end{figure}
\begin{table}[htbp]
  \caption{Fit parameters for effective LO potential with the $\pi\pi$-type source.}
  \vspace{2mm}
  \centering
  \scalebox{0.9}{
  \begin{tabular}{cccccccccc|c}
     & $a_0$ & $a_1$ & $a_2$ & $a_3$ & $a_4$ & $a_5$ & $a_6$ & $a_7$ & $a_8$ & $\chi^2 / {\rm dof.}$ \\ \hline \hline
    Fit & -0.0821(42)& 8.04(38)& 5.15(44)& -5.94(11)& -0.649(93)& 3.995(74)& 0.548(21)& 4.670(15)& 2.009(26) & 0.85\\
    Fit$^+$ &-0.0976(75)& 6.95(63)& 5.75(56)& -5.20(10)& -0.09(10)& 3.658(99)& 0.525(34)& 4.611(17)& 2.0286(34) & 0.24\\
    Fit$^-$ & -0.0983(10)& 6.84(79)& 5.84(57)& -5.66(14)& -0.28(14)& 3.76(17)& 0.574(71)& 4.517(43)& 2.109(53) & 0.14
  \end{tabular}
  }
  \label{tab:LOpotential_pipi_fitparams}
\end{table}
\begin{table}[htbp]
  \caption{Fit parameters for effective LO potential with the $\rho$-type source.}
  \vspace{2mm}
  \centering
  \scalebox{0.9}{
  \begin{tabular}{cccccccccc|c}
     & $a_0$ & $a_1$ & $a_2$ & $a_3$ & $a_4$ & $a_5$ & $a_6$ & $a_7$ & $a_8$ & $\chi^2 / {\rm dof.}$ \\ \hline \hline
    Fit &  -0.1146(74) & 9.64(33) & 5.23(47) & -6.35(17) & -1.03(15) & 4.456(95) & 0.711(25) & 4.818(23) & 2.068(29) & 1.15\\
    Fit$^+$ &  -0.124(11) & 8.89(65) & 5.83(68)& -5.57(16)& -0.42(15)& 4.12(14)& 0.712(41)& 4.752(27)& 2.111(36) & 0.29\\
    Fit$^-$ &  -0.120(17)& 9.1(1.1)& 5.75(97)& -6.18(21)& -0.74(23)& 4.34(27)& 0.81(11)& 4.650(72)& 2.227(79) & 0.18
  \end{tabular}
  }
  \label{tab:LOpotential_rho_fitparams}
\end{table}

We fit LO potentials {with a sum of Gaussian terms given by}
\begin{equation} \label{eq:fitfunc_3G}
  V(r) = a_0 e^{-(r-a_1)^2/a_2^2} + a_3 e^{-(r-a_4)^2/a_5^2} + a_6 e^{-(r-a_7)^2/a_8^2}.
\end{equation}
For the fit, we {utilize data projected to the $l=1$ component} by the partial wave decomposition at $r = [2, 14.8]$ as already discussed,
combined with the original lattice data at $r \leq 2$, to which the partial wave decomposition cannot be reliably applied.
We also remove data at very short distances ($r = 0, 1$) since they suffer from large discretization errors.
Remaining systematic uncertainties caused by non-smoothness at short distances are estimated by differences among three different fit results:
a result using all allowed data (Fit),
a result removing data at $r \le 0.32$ fm which are significantly larger than Fit (Fit$^-$),
and a result removing data at $r \le 0.32$ fm which are significantly smaller than Fit (Fit$^+$).
The fit results using all allowed data (Fit) are shown in Figure~\ref{fig:LOpotentials_fitresult}, and comparisons of three fit results at short distances are given in Figure~\ref{fig:LOpotentials_fitsys}.
Resultant fit parameters and $\chi^2 / {\rm dof}$ are given in Table.~\ref{tab:LOpotential_pipi_fitparams} and \ref{tab:LOpotential_rho_fitparams}.
As seen in Fig.\ref{fig:LOpotentials_fitsys},
the non-smooth behavior of the potential at short distances,
which is probably caused by contaminations from higher partial waves,
affects the fit result at $r \lesssim 0.25$ fm.
Since the removal of such contaminations at short distances is impractical,
we estimate systematic errors for physical observables by differences among fit results, taking the result using all allowed data as a central value.

\begin{figure}[tbp]
  \includegraphics[width=0.6\hsize,clip]{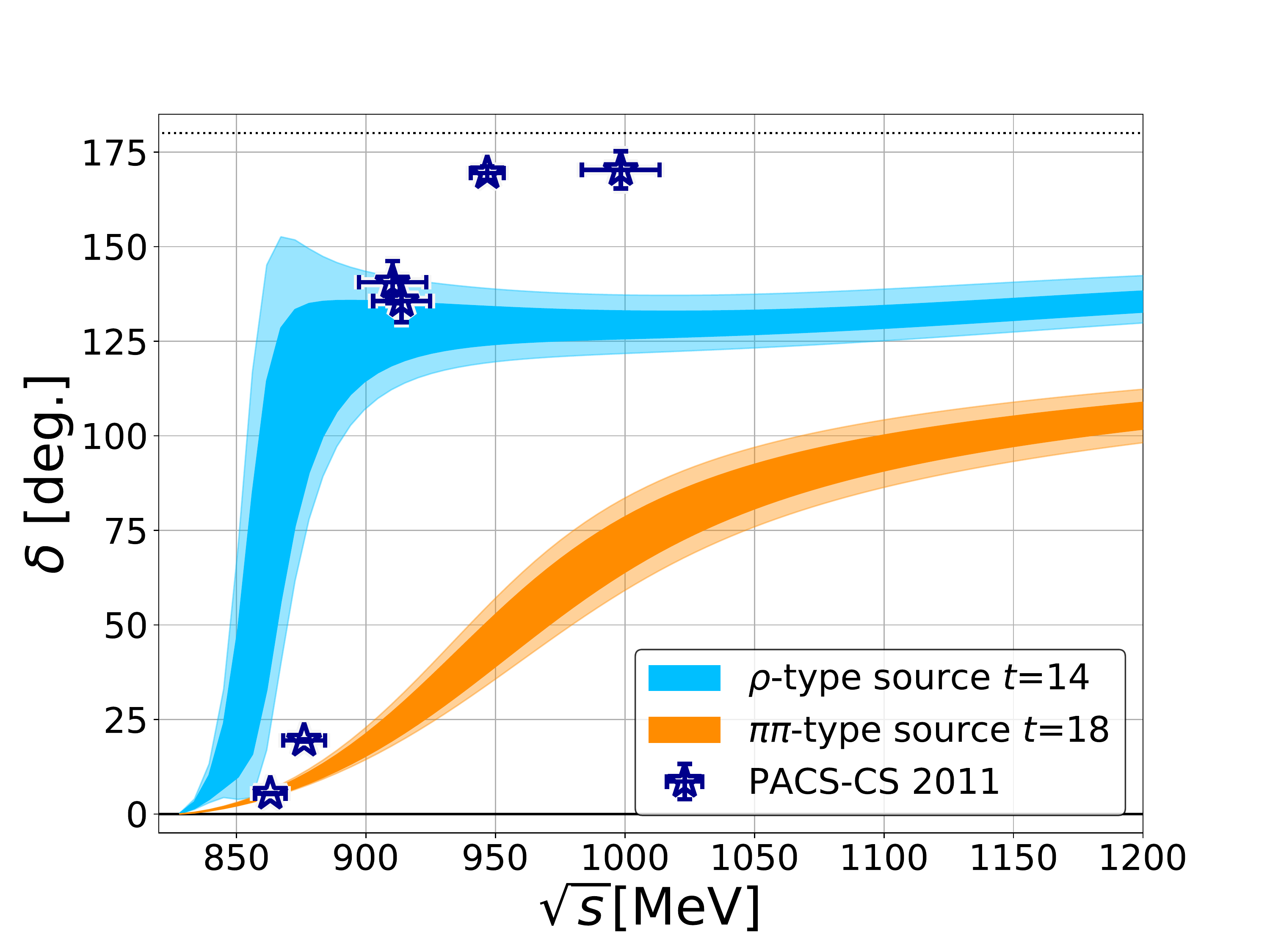}
  \caption{Phase shifts at the LO {analysis}.
    Blue (orange) band shows the $\rho$-type ($\pi\pi$-type) source result.
    Statistical errors are given by dark color bands, whereas systematic errors
  estimated by three different fits at short distances are represented by light color bands.
  The previous finite-volume results by the PACS-CS Collaboration~\cite{Aoki:2011yj} are
  also given by navy stars for comparison.}
  \label{fig:LOphaseshift}
\end{figure}
Figure~\ref{fig:LOphaseshift} shows phase shifts obtained from fitted potentials,
where systematic errors associated with removals of data at short distances are shown by light color bands on top of statistical errors by dark color bands.
Shown together is the previous finite-volume result reported in Ref.~\cite{Aoki:2011yj}, which employs the same gauge configurations.
The phase shift obtained with the $\rho$-type source crosses 90 degrees around $\sqrt{s} \approx 870$ MeV,
while it only reaches around 130 degrees as the energy increases.
On the other hand,
the phase shift obtained with the $\pi\pi$-type source crosses 90 degrees
at much higher energy, around $\sqrt{s} \approx 1050$ MeV,
with much broader width.
These behaviors are probably caused by truncation errors of the derivative expansion for the LO potential.
Since the $\rho$-type source strongly overlaps the $\rho$ resonance state, which corresponds to the ground state in this setup, the resultant phase shift with the $\rho$ source reproduces the $\rho$ resonance structure relatively well.
On the other hand, since the $\pi\pi$-type source mainly overlaps P-wave $\pi\pi$ scattering states, which appear in the energy region far above the $\rho$ resonance in this lattice setup,
it is difficult for the phase shift with the $\pi\pi$-type source to capture the resonance structure correctly.

%% N$^2$LO analysis %%
\subsection{The N$^2$LO analysis}
\begin{figure}[tbp]
  \hspace{-10mm}
  \begin{tabular}{cc}
  \begin{minipage}{0.5\hsize}
    \includegraphics[width=80mm,clip]{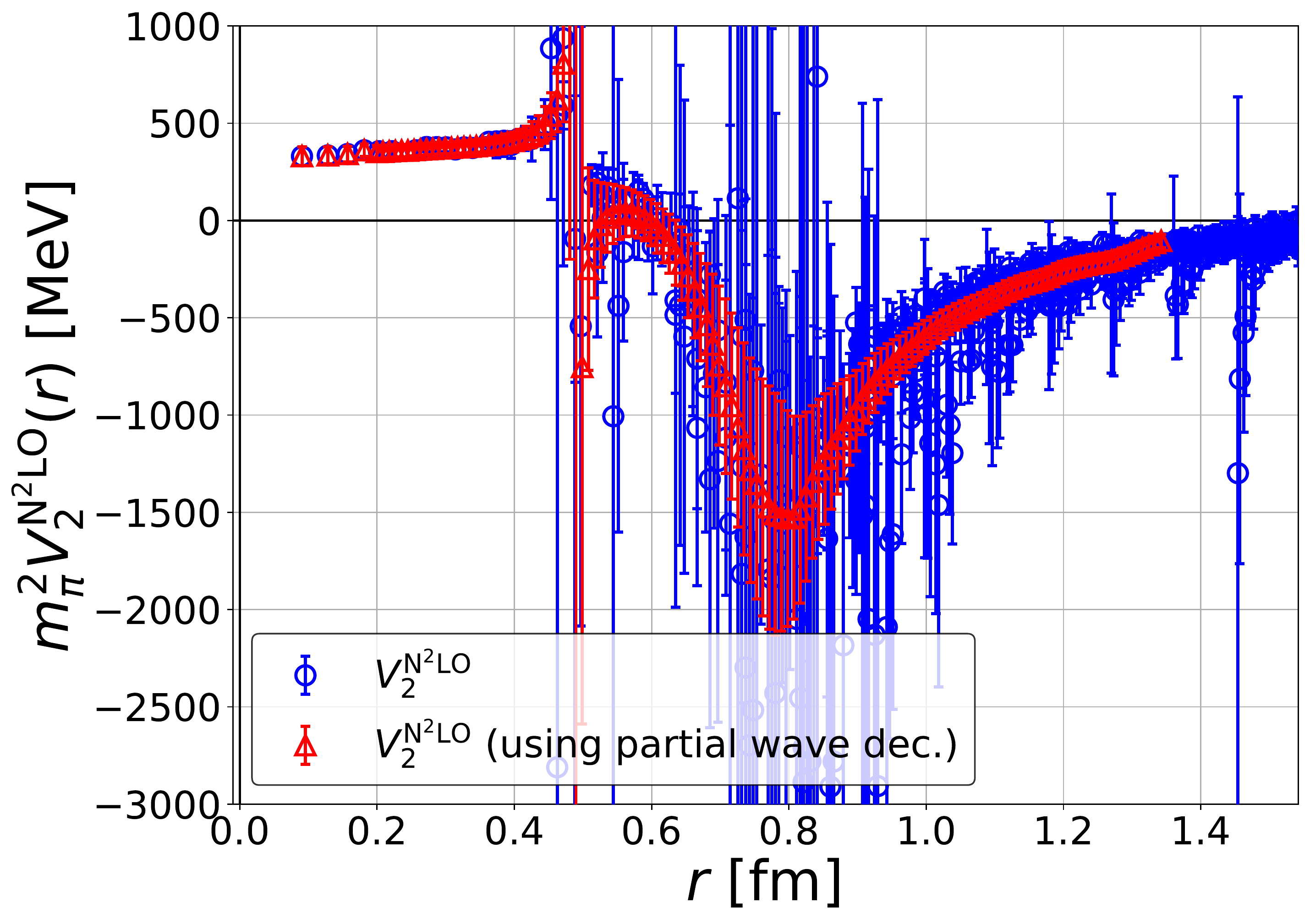}
  \end{minipage} &
  \begin{minipage}{0.5\hsize}
    \includegraphics[width=80mm,clip]{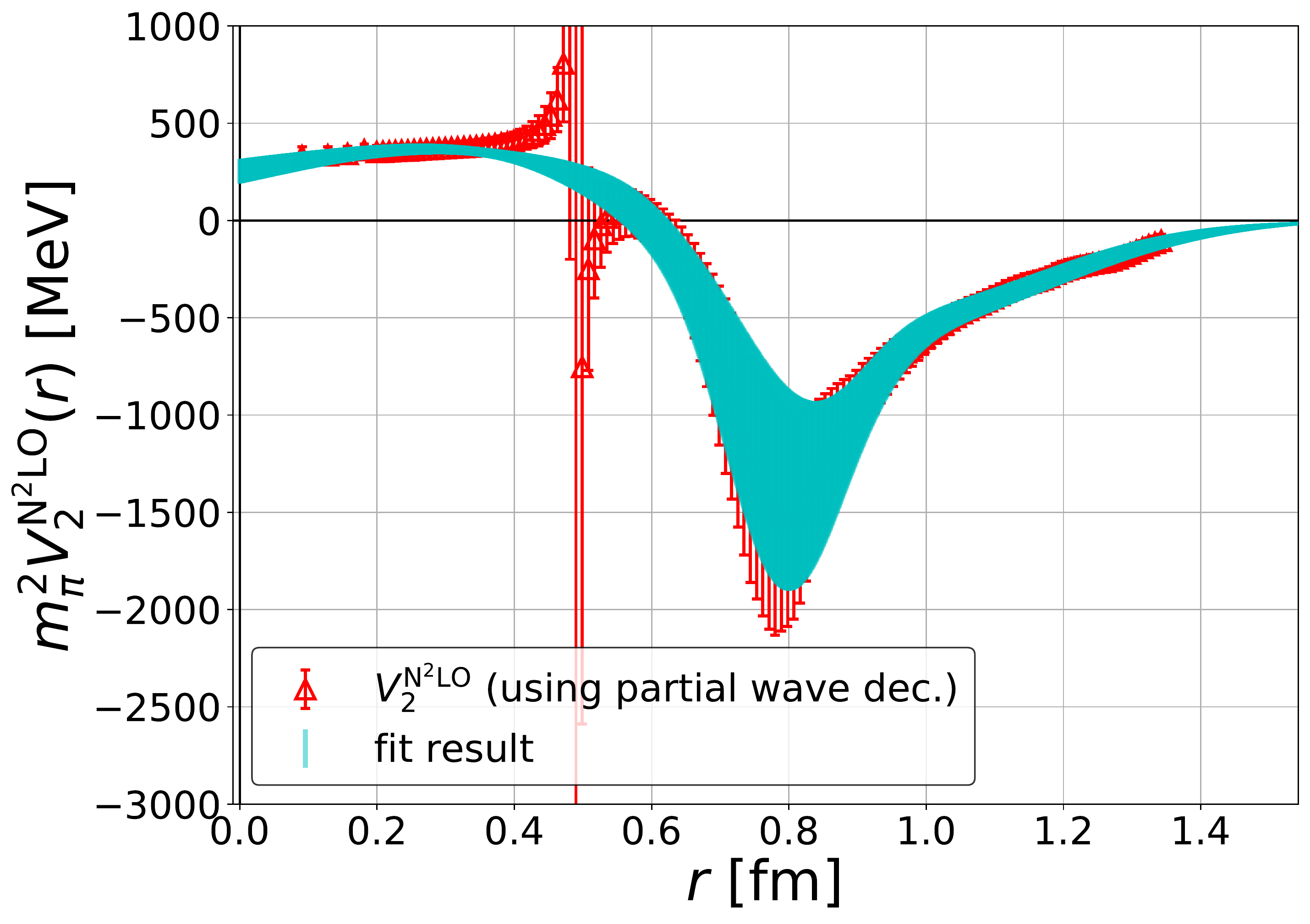}
  \end{minipage} \\
  \begin{minipage}{0.5\hsize}
    \includegraphics[width=80mm,clip]{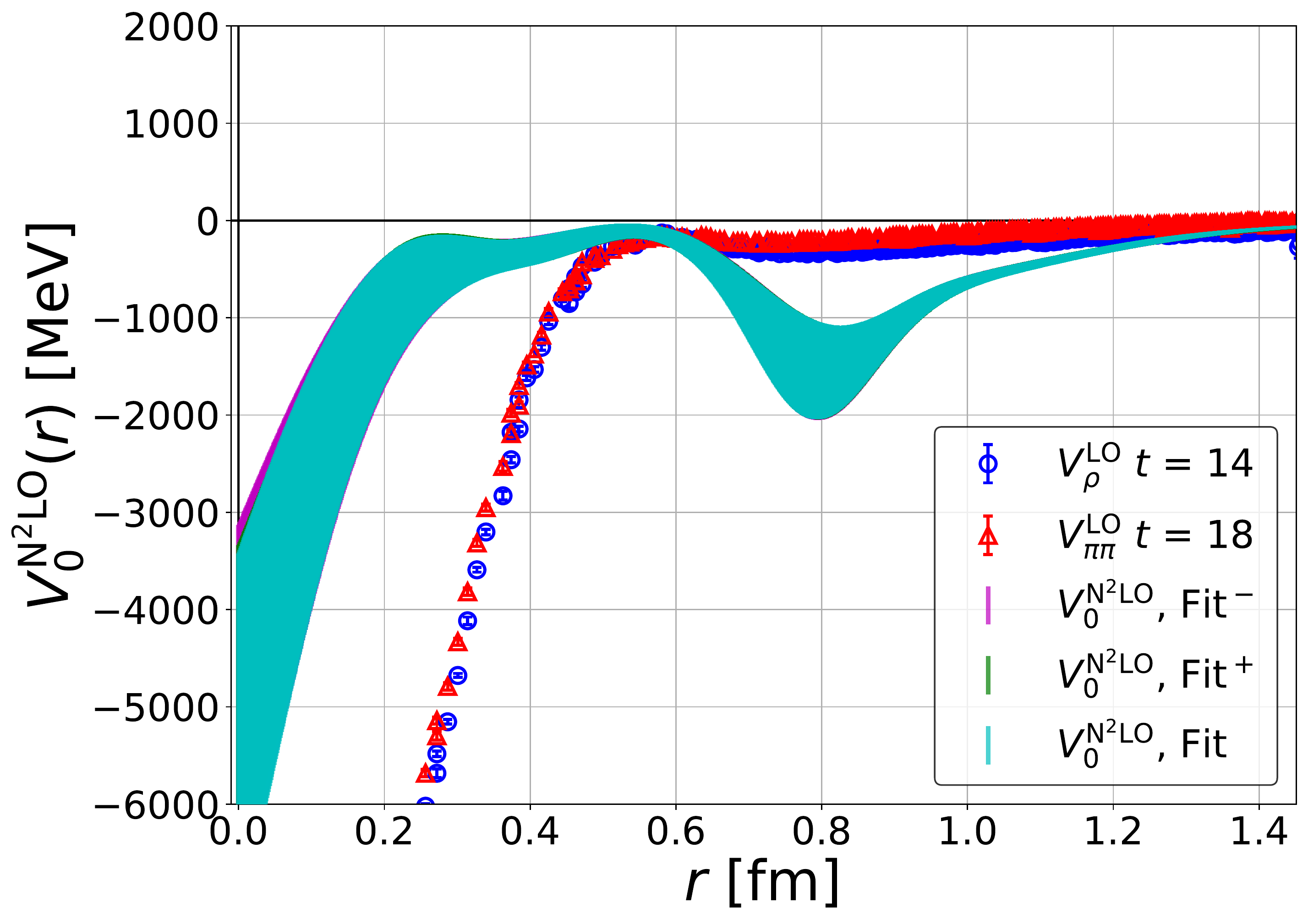}
  \end{minipage}
\end{tabular}
\caption{Effective N$^2$LO potentials. (Upper left) $V_2^{\rm N^2LO}$ determined from raw data (blue circles) and data obtained with the partial wave decomposition (red triangles). (Upper right) The fit result (cyan band) using the decomposed data (red triangles).
(Lower) $V_0^{\rm N^2LO}$ obtained by three fit results
%, Fit(green), Fit$^-$(light greem) and Fit$^+$(dark green).
, Fit(cyan), Fit$^-$(magenta) and Fit$^+$(green).
Shown together are the effective LO potentials for a comparison. }
\label{fig:NLOpotentials}
\end{figure}

% details of extraction of N$^2$LO potentials
As we have two LO potentials, we can proceed to the N$^2$LO analysis.
The effective N$^2$LO potentials are obtained through eq.(\ref{eq:NLOlineqs}) as
\begin{eqnarray}
  V_2^{\rm N^2LO}(r) &=& \frac{V_{\rho}^{\rm LO}(r) - V_{\pi \pi}^{\rm LO}(r)}{\nabla^2 R_{\rho}(r)/R_{\rho}(r) - \nabla^2 R_{\pi\pi}(r)/R_{\pi\pi}(r)} \label{eq:solutions_NLOlineqs_V2} \\
  V_0^{\rm N^2LO}(r) &=& V_{\rho}^{\rm LO}(r) - V_2^{\rm N^2LO}(r) \nabla^2 R_{\rho}(r)/R_{\rho}(r). \label{eq:solutions_NLOlineqs_V0}
\end{eqnarray}

In Fig.~\ref{fig:NLOpotentials} (upper left), we show $V_2^{\rm N^2LO}$ obtained from raw data (blue points),
and {$l=1$ data with the partial wave decomposition} (red points).
Thanks to the removal of higher partial wave contaminations,
we can significantly reduce fluctuations of $V_2^{\rm N^2LO}$, as seen in the figure.

A somewhat singular behavior at $r \approx 0.5$ fm is caused by a vanishing
denominator of $V_2^{\rm N^2LO}$ in Eq.(\ref{eq:solutions_NLOlineqs_V2}).
We however expect that this singular behavior is canceled by a vanishing numerator at the same point.
As discussed in Appendix~\ref{appex:fitassumption},
this expectation is shown to be true as long as the N$^4$LO (and higher order) terms in the derivative expansion are negligible.
Furthermore, we assume that $1-2 \mu V_2^{\rm N^2LO} > 0$, which is also shown to be true
in Appendix~\ref{appex:fitassumption} if the N$^4$LO or higher order terms are negligible.
We thus fit $V_2^{\rm N^2LO}$ (red point) by a smooth function,
a 3-Gaussian function in Eq~(\ref{eq:fitfunc_3G}), where data with $1-2 \mu V_2^{\rm N^2LO} \le 0$
are excluded in the fit.
Fit parameters for $V_2^{\rm N^2LO}$ are summarized in Table~\ref{tab:NLO_fitparams_V2NLO} and the fit result is shown by a cyan band in Fig.~\ref{fig:NLOpotentials}~(upper right).
Since significant non-smooth behavior is not observed for $V_2^{\rm N^2LO}$ at short distances,
 systematic errors associated with removals of data mentioned before are not included in the analysis for $V_2^{\rm N^2LO}$.
\begin{table}[htbp]
  \caption{Fit parameters of the $V_2^{\rm N^2LO}$.}
  \vspace{2mm}
  \centering
  \scalebox{1.0}{
  \begin{tabular}{ccccccccc|c}
     $a_0$ & $a_1$ & $a_2$ & $a_3$ & $a_4$ & $a_5$ & $a_6$ & $a_7$ & $a_8$ & $\chi^2 / dof$  \\ \hline \hline
     -12.8(6.4)& 8.82(24)& 1.37(11)& -9.4(5.8)& 9.86(94)& 3.97(78)& 5.7(4.7)& 4.8(2.4)& 6.48(72) & 0.063\\
  \end{tabular}
  }
  \label{tab:NLO_fitparams_V2NLO}
\end{table}
\begin{table}[htbp]
  \caption{Fit parameters of the Laplacian term $\nabla^2 R_{\rho} / R_{\rho}$.}
  \vspace{2mm}
  \centering
  \scalebox{0.9}{
  \begin{tabular}{cccccccccc|c}
     & $a_0$ & $a_1$ & $a_2$ & $a_3$ & $a_4$ & $a_5$ & $a_6$ & $a_7$ & $a_8$ & $\chi^2 / dof$  \\ \hline \hline
    Fit & -0.0271(12)& 9.04(43)& 6.21(36)& -1.135(25)& -0.70(13)& 4.33(12)& 0.1452(78)& 4.743(26)& 2.143(36) & 3.44\\
    Fit$^+$ & -0.0270(18)& 9.03(67)& 6.25(52)& -0.993(19)& -0.07(11)& 4.17(14)& 0.175(18)& 4.591(46)& 2.264(61) & 1.54\\
    Fit$^-$ & -0.0249(16)& 9.71(61)& 5.76(50)& -1.165(24)& -0.71(15)& 4.56(15)& 0.188(20)& 4.550(60)& 2.348(69) & 1.09\\
  \end{tabular}
  }
  \label{tab:NLO_fitparams_rholap}
\end{table}

Let us consider a determination of $V_0^{\rm N^2LO}(r)$ next.
We first fit the Laplacian term $\nabla^2 R_{\rho}(r)/R_{\rho}(r)$ by the same 3-Gaussian function, and resultant parameters are given in Table~\ref{tab:NLO_fitparams_rholap}.
We then obtain $V_0^{\rm N^2LO}(r)$ by combining all the fit results in Eq.~(\ref{eq:solutions_NLOlineqs_V0}).
We estimate systematic errors of $V_0^{\rm N^2LO}(r)$ at short distances through those of $V_{\rho}^{\rm LO}$ and $\nabla^2 R_{\rho}(r)/R_{\rho}(r)$.
Figure~\ref{fig:NLOpotentials}(lower) shows the resultant $V_0^{\rm N^2LO}$, together with effective LO potentials, $V_{\rho}^{\rm LO}$ {and $V_{\pi\pi}^{\rm LO}$}, for a comparison.
As expected, there exists a large difference between $V_{\rho,\pi\pi}^{\rm LO}$ and $V_0^{\rm N^2LO}$ in Fig.~\ref{fig:NLOpotentials}~(lower).

\begin{figure}[htbp]
  \begin{tabular}{cc}
  \begin{minipage}{0.5\hsize}
    \includegraphics[width=80mm,clip]{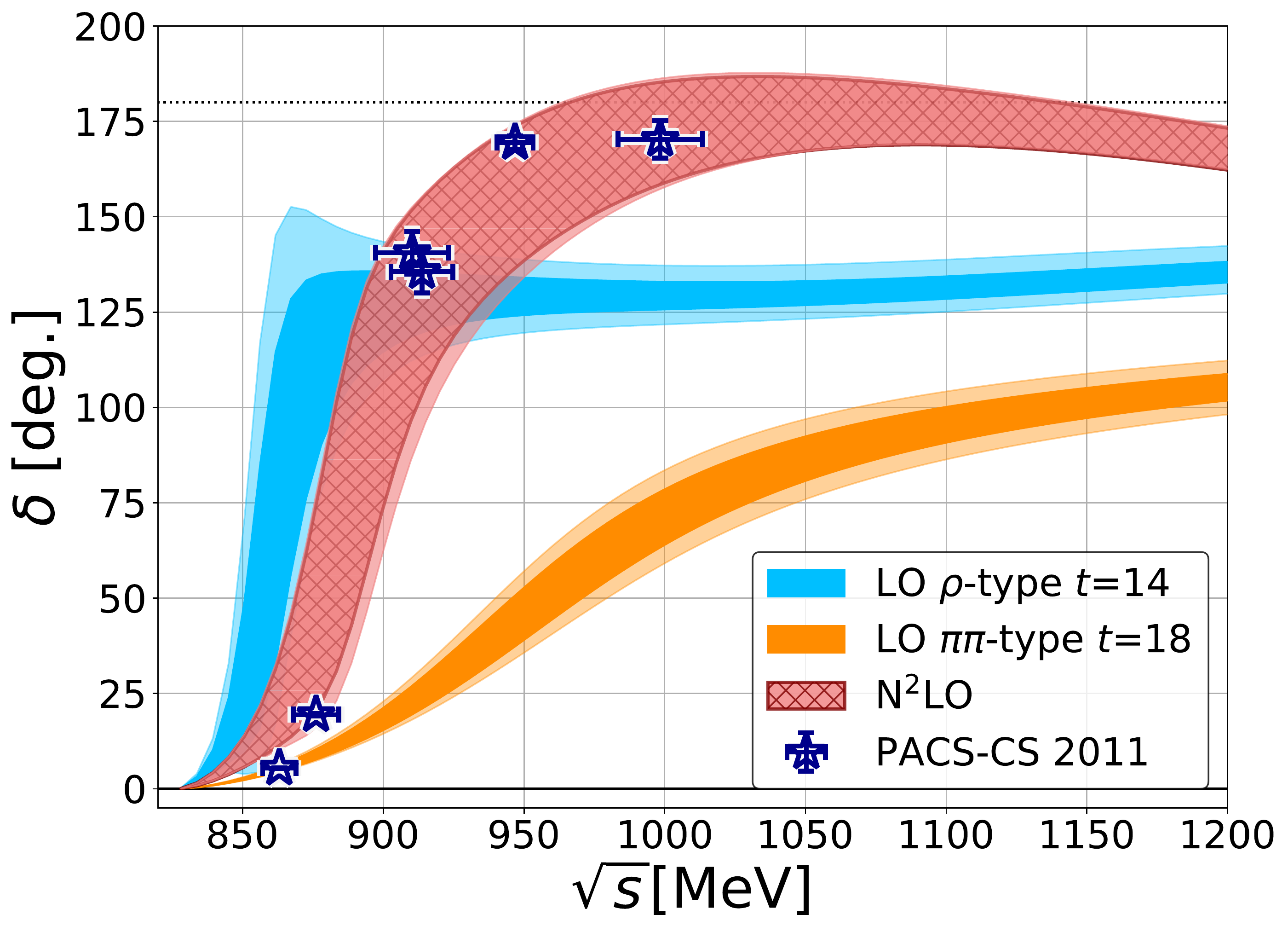}
  \end{minipage} &
  \begin{minipage}{0.5\hsize}
    \includegraphics[width=80mm,clip]{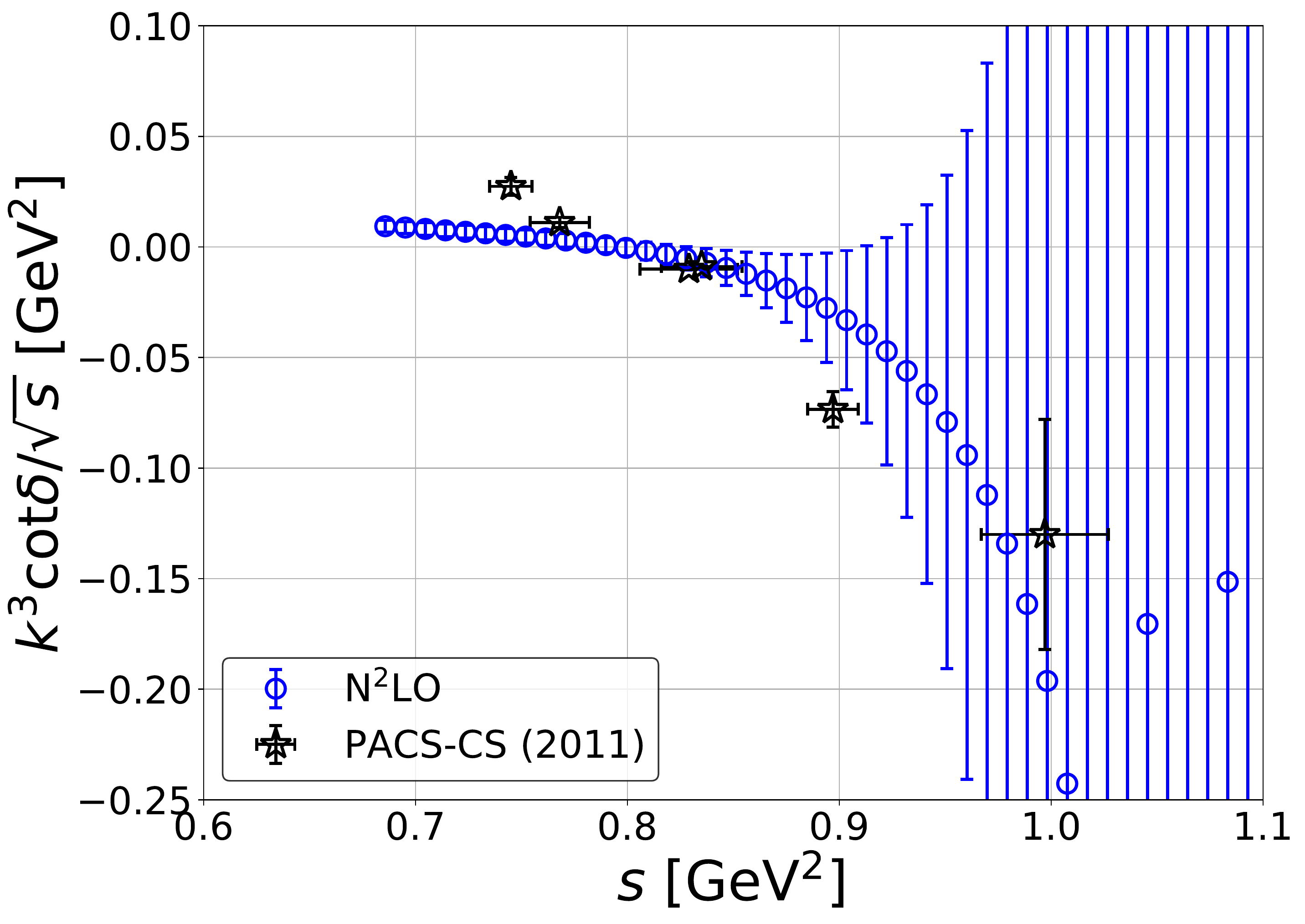}
  \end{minipage}
\end{tabular}
\caption{The N$^2$LO phase shifts (left) and $k^3 \cot \delta_1 / \sqrt{s}$ (right),
together with LO results (left figure) and previous finite-volume result by the PACS-CS Collaboration~\cite{Aoki:2011yj} (both figure) for comparisons.
Large statistical errors at $s > 0.9$ GeV$^2$ in $k^3 \cot \delta_1 / \sqrt{s}$ (right) are mainly caused by a divergent behavior of $\cot \delta$ at the phase shift around 180 degrees.
}
\label{fig:NLOphaseshiftandk3cotd}
\end{figure}

% NLO phase shift and k3cotd/sqrts
To obtain the N$^2$LO phase shifts, we solve the radial Schr\"odinger equation with the N$^2$LO potential, rewritten as
\begin{equation}
  \left( \frac{d^2}{dr^2} - \frac{l (l+1)}{r^2} - \frac{2\mu V_0(r) - k^2}{1 - 2\mu V_2(r)} \right) \phi = 0.
\end{equation}
The N$^2$LO phase shifts and the corresponding $k^3 \cot \delta_1 / \sqrt{s}$ are shown in Figure~\ref{fig:NLOphaseshiftandk3cotd}, together with the LO phase shifts and the previous finite-volume result for comparisons.
We have checked that the N$^2$LO phase shifts do not vary beyond the magnitude of statistical errors even if we choose a different timeslice for the $\pi\pi$-type source in the N$^2$LO analysis.

%% important discussion here. %%
As can be seen in Fig.~\ref{fig:NLOphaseshiftandk3cotd}, except for the region $s < 0.75 \ {\rm GeV}^2$ ($\sqrt{s} < 870$ MeV), the N$^2$LO phase shifts and $k^3 \cot \delta_1 / \sqrt{s}$ are roughly consistent with the finite-volume results.
The deviation observed in the low-energy region
can be understood from the truncation error of the derivative expansion as discussed in Sect.~\ref{sect:halqcdmethod}.
In this study, the calculations are performed only in the center-of-mass energy frame,
  where the corresponding energy levels on the current lattice volume
  do not cover the low-energy region near the $\pi\pi$ threshold.
  Therefore, the N$^2$LO approximation in this study could suffer from the large truncation error of the derivative expansion
  in such a low-energy region.
This discrepancy actually affects a determination of some resonance parameters as will be discussed later.
The detailed investigation is left for future studies since it needs much higher precision
with possibly an additional technical development of the laboratory-frame calculation\cite{Aoki:2020cwk}.

\subsection{Resonance parameters}
In this subsection, we extract resonance parameters for the $\rho$ meson in the N$^2$LO analysis
using two different methods.

\subsubsection{Breit-Wigner fit}
\begin{figure}[htbp]
  \centering
  \includegraphics[width=90mm,clip]{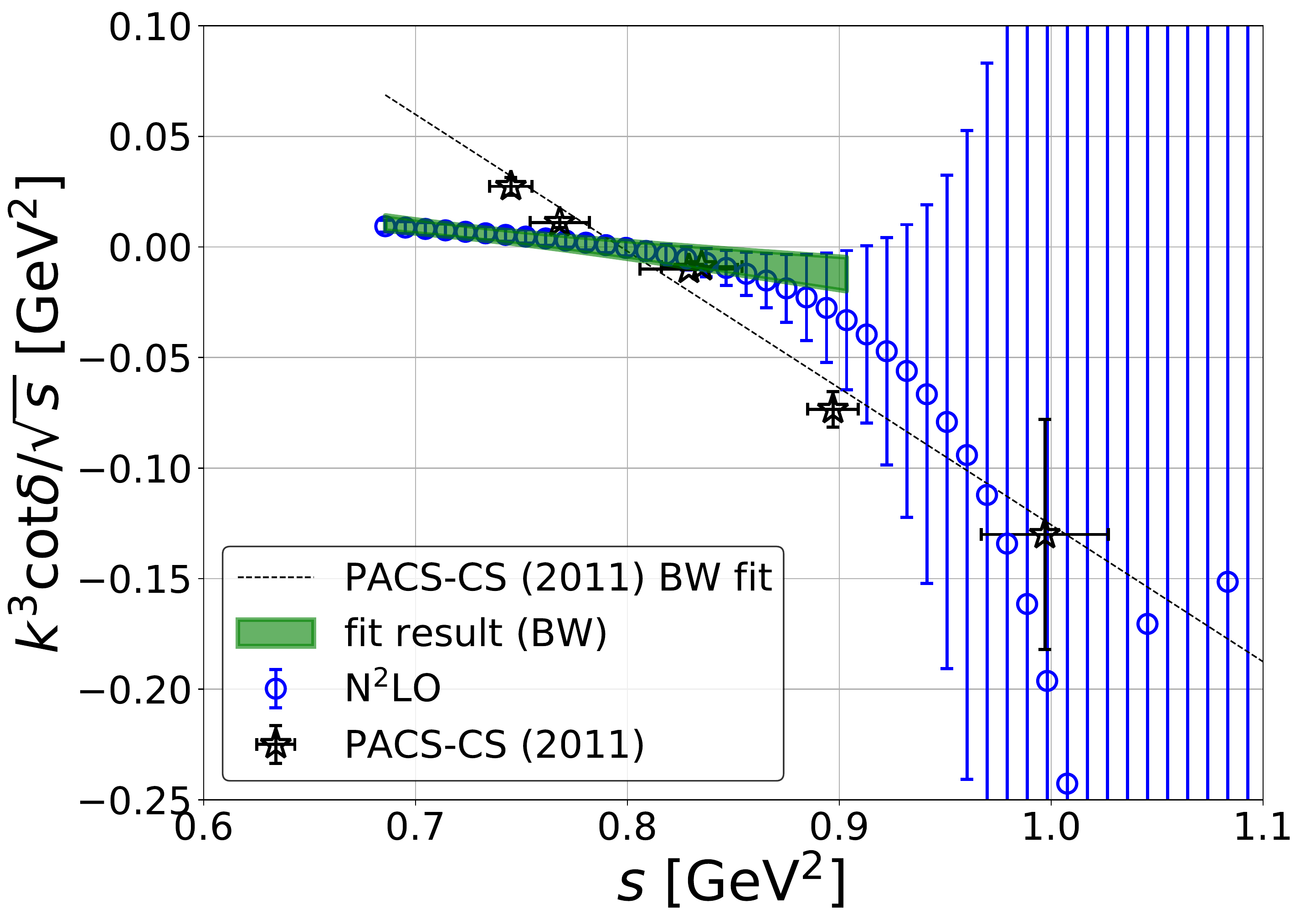}
  \caption{The Breit-Wigner fit for the N$^2$LO phase shifts $k^3 \cot \delta_1 (k) / \sqrt{s}$ (blue points).
    The green band represents the fit with statistical errors and a range of the energy used in the fit. We also show data and Breit-Wigner fit of PACS-CS (2011)~\cite{Aoki:2011yj} by the black points and the dashed line, respectively, for a comparison. }
  \label{fig:BWfitresults}
\end{figure}

We first extract resonance parameters in the conventional way,
by fitting the scattering phase shifts
with the Breit-Wigner form as
\begin{eqnarray} \label{eq:BWfit}
  \frac{k^3 \cot \delta_1 (k)}{\sqrt{s}} &=& \frac{6 \pi}{g_{\rho \pi \pi}^2} (m_{\rho}^2 - s),
\end{eqnarray}
where $m_{\rho}$ and $g_{\rho \pi \pi}$ are fit parameters
corresponding to a resonance mass and a $\rho\to\pi\pi$ effective coupling, respectively.
We show the fit result in Fig.~\ref{fig:BWfitresults}, which gives
\begin{eqnarray}
  &&m_{\rho} = 888 (19) (^{+6}_{-2}) \ {\rm MeV},\\
  &&g_{\rho \pi \pi} = 13.4 (2.6) (^{+0.8}_{-0.0}),
\end{eqnarray}
with $\chi^2 / {\rm dof} = 0.18$, where
the first errors are statistical and the second ones are systematic errors
associated with the short-range behavior of $V_0^{\rm N^2LO}$.

We have checked that resonance parameters remain unchanged within statistical errors
even if we add a centrifugal barrier modification as a higher order term in $k^2$\cite{VonHippel:1972fg} to the standard Breit-Wigner form in Eq.~(\ref{eq:BWfit}).

\subsubsection{Direct pole search}
Theoretically, a resonance state is defined as a pole of the S-matrix on the second Riemann sheet,
which provides us the second method to extract resonance parameters in the HAL QCD method.
To access the S-matrix in complex energy region, we solve the Schr\"odinger equation
with arguments rotated by
$r \to r e^{i \theta}, k \to k e^{-i \theta}$\cite{Giraud:2005gz,Aguilar:1971ve,Balslev:1971vb},
which reads
\begin{equation}
  \left( \frac{d^2}{dr^2} - \frac{l (l+1)}{r^2} - \frac{2\mu e^{2 i \theta}V_0(e^{i \theta}r) - k^2}{1 - 2\mu V_2(e^{i \theta}r)} \right) \phi = 0.
\end{equation}
The regular solution $\phi$ {to} this equation behaves at long distances as
\begin{equation}
  \phi \to \frac{i}{2} \left[ {\mathcal J}_l(ke^{-i \theta}) \hat h_l^{-}(kr) - {\mathcal J}^{*}_l(ke^{-i \theta}) \hat h_l^{+}(kr) \right],
\end{equation}
where $\hat h_l^{\pm}(z) = \hat n_l(z) \pm i \hat j_l (z)$ are the Riccati-Hankel functions and ${\mathcal J}_l$ is the Jost function {for} the angular momentum $l$.
The S-matrix on the ray of $ke^{-i \theta}$ can therefore be obtained as
\begin{equation}
  s_l(ke^{-i \theta}) = \frac{{\mathcal J}^*_l(ke^{-i \theta})}{{\mathcal J}_l(ke^{-i \theta})},
\end{equation}
from which we can search a pole position $k_{\rm pole} = |k_{\rm pole}|e^{-i \theta_{\rm pole}}$ by changing an input $\theta$ and $k$.
The resonance mass and the decay  width are extracted from the pole position $\sqrt{s}_{\rm pole}$ as
\begin{equation}
  \sqrt{s}_{\rm pole} = 2 \sqrt{k_{\rm pole}^2 + m_{\pi}^2} = m_{\rho} - i \Gamma_{\rho} / 2,
\end{equation}
where the decay width is related to the coupling constant $g_{\rho \pi \pi}$ as
\begin{equation}
  g_{\rho \pi \pi} = \sqrt{\frac{6 \pi \Gamma_{\rho} m_{\rho}^2}{k_{\rho}^3}},
  \quad  k_{\rho} := \sqrt{ m_{\rho}^2 / 4 - m_{\pi}^2 }.
\end{equation}

The direct pole search gives
\begin{eqnarray}
  m_{\rho} &=& 886 (17) (^{+4}_{-1}) \ {\rm MeV},\\
  \Gamma_{\rho}/2 &=& 22 (8.6) (^{+4.5}_{-0.0}) \ {\rm MeV}, \\
  g_{\rho \pi \pi} &=& 12.7 (2.9) (^{+0.7}_{-0.0}),
\end{eqnarray}
where the first errors are statistical while the second ones are systematic errors
associated with the short-range behavior of $V_0^{\rm N^2LO}$.

\subsubsection{Comparison to the previous result}

\begin{figure}[tbp]
  \centering
  \includegraphics[width=0.5\hsize,clip]{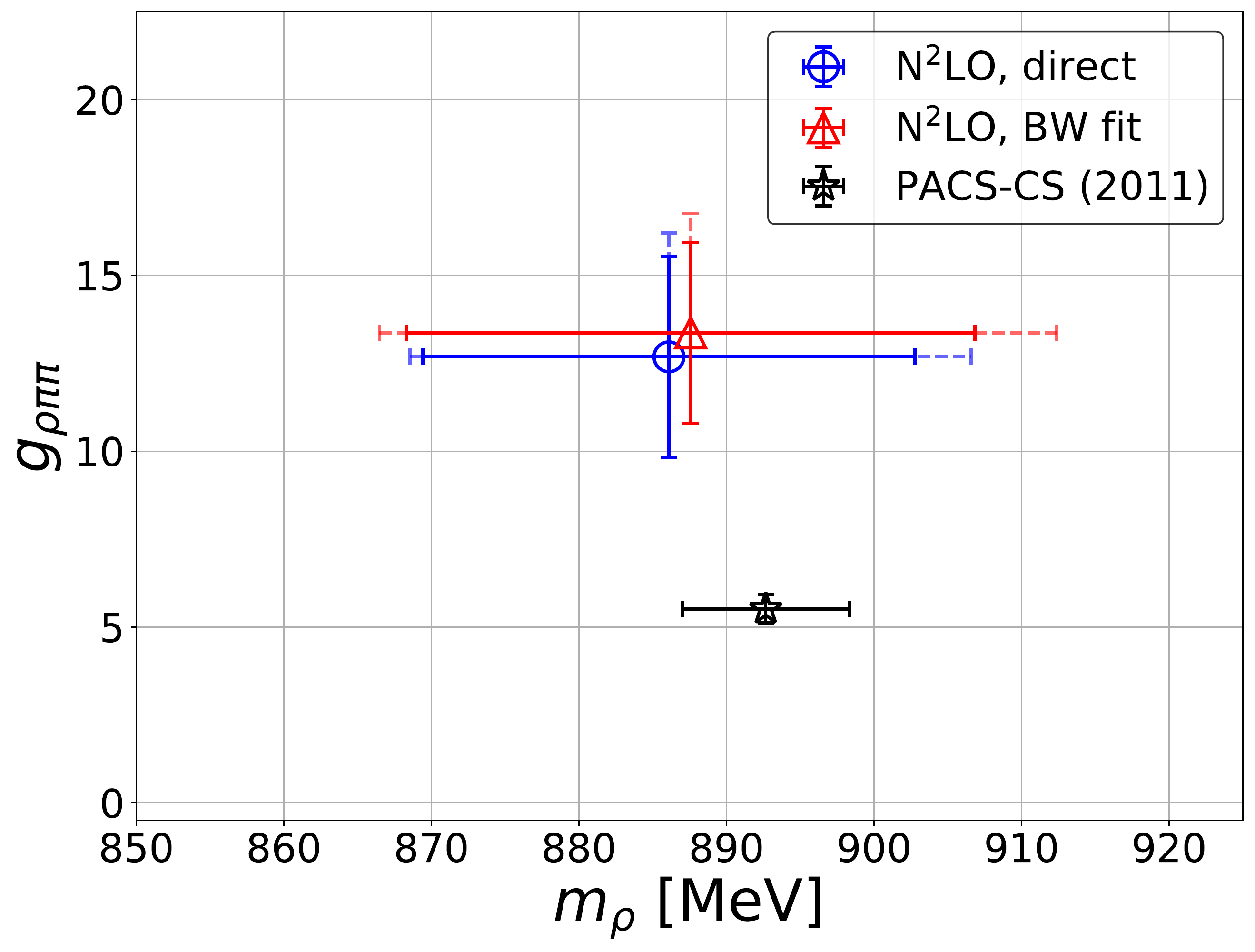}
  \caption{A comparison of N$^2$LO resonance parameters from the direct pole search (blue circle) and the Breit-Wigner fit (red triangle).
    Vertical and horizontal axes represent the coupling $g_{\rho \pi\pi}$ and {the mass} $m_{\rho}$, respectively.
    Statistical (systematic) errors are represented by solid (dashed) bars.
    Shown together is the result from PACS-CS (2011)~\cite{Aoki:2011yj} (black star).
    Errors for PACS-CS (2011) are statistical only.
    }
  \label{fig:resparams_summary}
\end{figure}

Let us compare our N$^2$LO results with the previous PACS-CS (2011) result using the finite-volume method~\cite{Aoki:2011yj}, both of which employ the same gauge configurations.
We plot $m_{\rho}$ and $g_{\rho \pi \pi}$ in Fig.~\ref{fig:resparams_summary}.
While $m_{\rho}$'s are consistent with each other in all three cases,
the Breit-Wigner fit, the pole search and  the PACS-CS (2011) result,
coupling constants in our both results are about twice as large as the previous one.
This discrepancy can be clearly seen as a difference in slopes of $k^3 \cot \delta_1 / \sqrt{s}$ data at $s < 0.9$ GeV$^2$ in Fig.~\ref{fig:BWfitresults},
which directly correspond to the coupling as $- 6 \pi / g_{\rho \pi \pi}^2$.
In particular, a significant disagreement
between the lowest energy level in PACS-CS and our data at $s \approx 0.75$ GeV$^2$
is a main source of the discrepancy for the slope.
We note that the lowest energy level in PACS-CS (2011) is obtained in the laboratory frame with ${\bf P}=(0,0,2\pi/L)$.
Since such a low-energy region cannot be covered by the center-of-mass frame employed in this study
and the non-locality of the potential in $I=1$ $\pi\pi$ systems turns out to be large,
%it is likely that our N$^2$LO approximation suffers from large truncation errors in the derivative expansion
our N$^2$LO approximation is likely to suffer from large truncation errors in the derivative expansion
at low energies.

This observation gives us a useful lesson
for the study of P-wave (or higher partial wave) resonances by the HAL QCD method
with the center-of-mass frame.
If the non-locality of the potential happens to be large,
the truncation errors could be large at low-energies near the threshold.
While a resonance mass is likely to be well reproduced as long as
the resonance appears
in the energy region accessible in the center-of-mass frame,
the decay width (the effective coupling) may suffer from larger systematics
since it is sensitive to energy dependence on a much wider range around the resonance.
As a possible option to control this systematics,
if the resonance mass can be roughly guessed,
one may tune lattice parameters such as a box size carefully so as to cover
a wide energy range even in the center-of-mass frame.
This procedure, however, is difficult in practice
and applicability for searches of unknown resonances is also limited.
The second option
is to establish the existence of a resonance
and to estimate its mass
by the HAL QCD method
in the center-of-mass frame, which is supplemented by the finite-volume method
in the laboratory frame to estimate its width reliably.
The third option is
to perform the HAL QCD method with
a combination of both the center-of-mass and laboratory frames.
In fact, a theoretical framework has been already proposed
for the HAL QCD method in the laboratory frame\cite{Aoki:2020cwk}.
While extraction of HAL QCD potentials from NBS wave functions in the laboratory frame
is indeed a numerical challenge, the first numerical trial is now ongoing.

%%%%% summary %%%%%
\section{Summary} \label{sect:summary}
We study the $I=1$ $\pi\pi$ interaction at $m_{\pi} \approx 411$ MeV, where the $\rho$ meson emerges as the resonance with $m_{\rho} \approx 892$ MeV.
We calculate all-to-all propagators by a combination of the one-end trick, the sequential propagator, and the covariant approximation averaging.
Thanks to those techniques, we successfully determine the potential in this channel at the N$^2$LO of the derivative expansion for the first time and calculate the resonance parameters of the $\rho$ meson.

The mass and decay width of the $\rho$ resonance are directly extracted
from the pole position of the S-matrix, $\sqrt{s}_{\rm pole} = 886(17) - i 22(9)$,
whose real part agrees with the $\rho$ resonance mass in the previous study but
whose imaginary part leads to the coupling constant $g_{\rho \pi \pi}$ twice as large as
the previous one.
Larger coupling $g_{\rho \pi \pi}$ originates from
the discrepancy in phase shifts at $s \lesssim 0.75$ GeV$^2$,
whose energy region cannot be covered in the center-of-mass frame of our lattice setup.
This observation provides a useful lesson for studies of P-wave resonances by the HAL QCD method
and future direction for the improvement is discussed.

Although the issue above remain to be verified explicitly, the result in this study
shows that hadronic resonances which require all-to-all calculations can be investigated with reasonable precisions
even at the N$^2$LO level in the HAL QCD method.
This study opens new doors toward understanding hadronic resonances by the HAL QCD method,
including more challenging systems such as $I=0$ $\pi \pi$ ($\sigma$ resonance),
$I=1/2$ $K \pi$ ($\kappa$ resonance) and exotic resonances.

\acknowledgements
The authors thank members of the HAL QCD Collaboration for fruitful discussions. We thank the PACS-CS Collaboration~\cite{Aoki:2008sm} and ILDG/JLDG~\cite{Amagasa:2015zwb} for providing their configurations. The numerical simulation in this study is performed on the HOKUSAI Big-Waterfall in RIKEN and the Oakforest-PACS in Joint Center for Advanced HighPerformance Computing (JCAHPC). The framework of our numerical code is based on Bridge++ code~\cite{Ueda:2014rya} and its optimized version for the Oakforest-PACS by Dr. I. Kanamori~\cite{Kanamori:2018hwh}.
This work is supported in part
by HPCI System Research Project (hp200108, hp210061),
by the Grant-in-Aid of the Japanese Ministry of Education, Sciences and Technology, Sports and Culture (MEXT) for Scientific Research (Nos.~JP16H03978,  JP18K03620, JP18H05236, JP18H05407, JP19K03847).
Y.A. is supported in part by the Japan Society for the Promotion of Science (JSPS).
S. A. and T. D. are also supported in part
by a priority issue (Elucidation of the fundamental laws and evolution of the universe) to be tackled by using Post ``K" Computer,
by Program for Promoting Researches on the Supercomputer Fugaku (Simulation for basic science: from fundamental laws of particles to creation of nuclei),
and by Joint Institute for Computational Fundamental Science (JICFuS).

The authors also thank the Yukawa Institute for Theoretical Physics (YITP) at Kyoto University.
Discussions during the YITP workshop YITP-X-19-03 on "Non-perturbative methods in quantum field theories and applications to elementary particle physics",
YITP-W-19-15 on "QUCS 2019" and
YITP-T-19-01 on "Frontiers in Lattice QCD and related topics" were useful to complete this work.

%%%%% appendix %%%%%
\appendix

\section{The one-end trick} \label{appex:oneend}
In this appendix, we briefly explain the one-end trick\cite{McNeile:2006bz}, which enables us to estimate a combination of two all-to-all propagators with a space summation by using a single noisy estimator.
Let us consider a combination of quark propagators given by
\begin{equation} \label{eq:oneend1}
  \sum_{\bf y} e^{i{\bf p \cdot y}} D_f^{-1}({\bf x}_1,t_1;{\bf y},t_0) \Gamma D_{f'}^{-1}({\bf y},t_0;{\bf x}_2,t_2),
\end{equation}
where $D_f^{-1}$ is a quark propagator with a flavor $f$, $\Gamma$ is some product of gamma matrices, and $x_i = ({\bf x}_i,t_i)$ are arbitrary.
We abbreviate color and spin indices for simplicity.
Such a structure typically appears at the source side of correlation functions including meson operators.
For example, in the separated diagram in Fig.~\ref{fig:diagrams}, it appears twice as
\begin{equation}
  \begin{split}
  (+) \sum_{{\bf y_1,y_2}} e^{i{\bf p_z \cdot y_1}} e^{-i{\bf p_z \cdot y_2}} &{\rm tr} \left[ \underline{ D^{-1} ({\bf x+r},t;{\bf y_1},t_0) \gamma_5 D^{-1} ({\bf y_1},t_0;{\bf x+r},t) }\gamma_5 \right] \\
  & \times {\rm tr} \left[ \underline{ D^{-1} ({\bf x},t;{\bf y_2},t_0) \gamma_5 D^{-1} ({\bf y_2},t_0;{\bf x},t) } \gamma_5 \right].
\end{split}
\end{equation}
The calculation of those structures naively needs two stochastic estimations for each, since each of them contains two all-to-all propagators.
The one-end trick, however, utilize the $\gamma_5$-Hermiticity of the Dirac operator to estimate that structure with a single noise insertion as follows.
\begin{equation}
  \begin{split}
    \sum_{\bf y} e^{i{\bf p \cdot y}} &D^{-1}(x_1;{\bf y},t_0) \Gamma D^{-1}({\bf y},t_0;x_2) \\
    &= \sum_{\bf y,z} e^{i{\bf p \cdot y}} D^{-1}(x_1;{\bf y},t_0) \delta_{\bf y,z} \Gamma D^{-1}({\bf z},t_0;x_2) \\
    &\approx \sum_{\bf y,z} e^{i{\bf p \cdot y}} D^{-1}(x_1;{\bf y},t_0) \left(\frac{1}{N_{\rm r}} \sum_{r=0}^{N_{\rm r}-1} \eta_{[r]}({\bf y}) \eta_{[r]}^{\dag}({\bf z}) \right) \Gamma D^{-1}({\bf z},t_0;x_2) \\
    &= \frac{1}{N_{\rm r}} \sum_{r=0}^{N_{\rm r}-1} \left( \sum_{\bf y} D^{-1}(x_1;{\bf y},t_0) \eta_{[r]}({\bf y}) e^{i{\bf p \cdot y}} \right)  \left( \sum_{\bf z} \gamma_5 D^{-1}(x_2;{\bf z},t_0) \gamma_5 \Gamma^{\dag} \eta_{[r]}({\bf z}) \right)^{\dag},
  \end{split}
\end{equation}
where we insert the stochastic estimator $\delta_{\bf y,z} \approx \frac{1}{N_{\rm r}} \sum_{r=0}^{N_{\rm r}-1} \eta_{[r]}({\bf z})  \eta_{[r]}^{\dag}({\bf y})$ in the second line and use the $\gamma_5$-Hermiticity in the last line.
We define ''one-end vectors`` as
\begin{eqnarray}
  \xi_{{\bf p},t_0[r]}(x) &\equiv& \sum_{\bf y} D^{-1}(x;{\bf y},t_0) \eta_{[r]}({\bf y}) e^{i{\bf p \cdot y}} \\
  \chi_{\Gamma,t_0[r]}(x) &\equiv& \sum_{\bf y} D^{-1}(x;{\bf y},t_0) \gamma_5 \Gamma^{\dag} \eta_{[r]}({\bf y}),
\end{eqnarray}
then  the final expression {becomes}
\begin{equation}
  \sum_{\bf y} D^{-1}({\bf x}_1,t_1;{\bf y},t_0) \Gamma D^{-1}({\bf y},t_0;{\bf x}_2,t_2) \approx \frac{1}{N_{\rm r}} \sum_{r=0}^{N_{\rm r}-1} \xi_{{\bf p},t_0[r]}({\bf x}_1,t_1) \otimes \chi_{\Gamma,t_0[r]}^{\dag} ({\bf x}_2,t_2) \gamma_5.
\end{equation}
The one-end vectors $\xi$ and $\chi$ are obtained by solving the linear equation $D \xi = \eta e^{i{\bf p \cdot y}}$ and $D \chi = \gamma_5 \Gamma^{\dag} \eta$, respectively. The dilution technique for noise reduction can be combined as well.
This trick is particularly suitable for the HAL QCD method since it does not introduce any stochastic estimations at the sink side, which
otherwise strongly affects spatial dependences of the NBS wave function.
Moreover, a numerical cost and stochastic noises are also reduced in accordance with a decrease in the number of noise vectors.

\section{Numerical evaluation for each diagram} \label{appex:diagrams}
\begin{figure}[htbp]
  \begin{tabular}{ccc}
  \begin{minipage}{0.3\hsize}
    \includegraphics[width=45mm,clip]{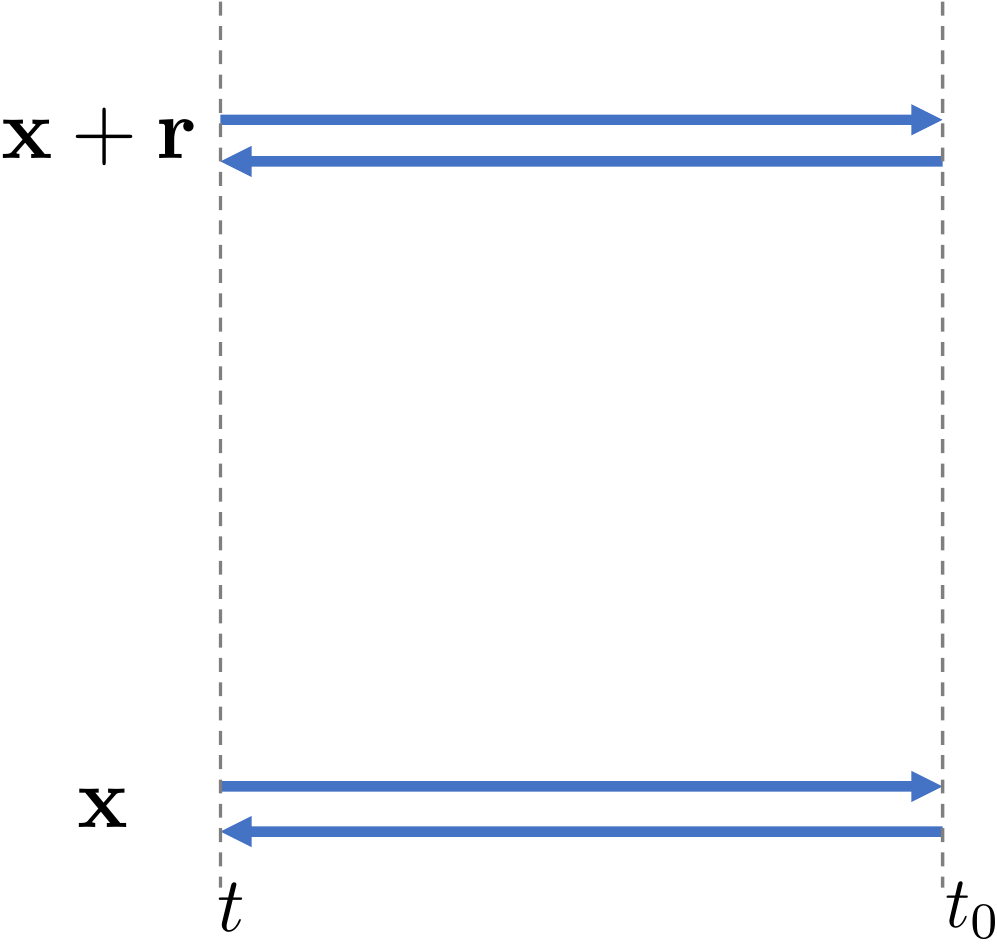}\\
    \centering
    separated diagram
  \end{minipage} &
  \begin{minipage}{0.3\hsize}
    \includegraphics[width=45mm,clip]{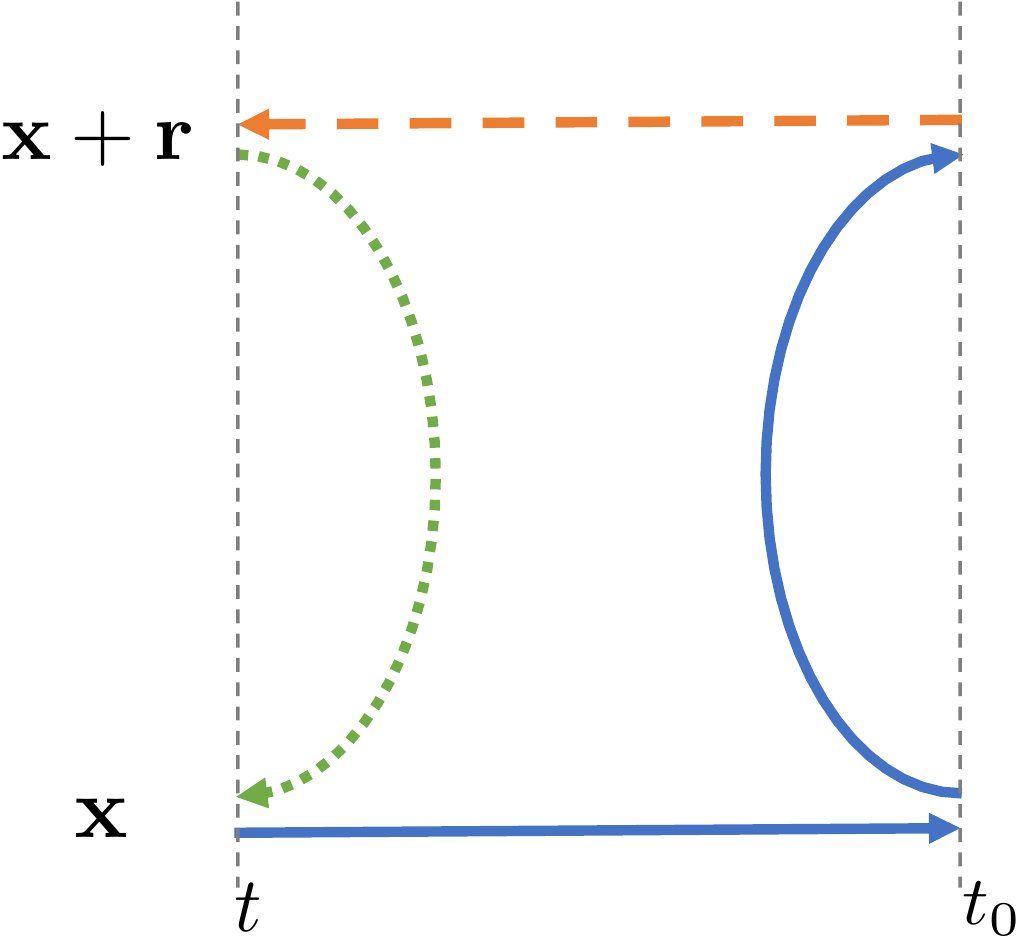}\\
    \centering
    box diagram
  \end{minipage} &
  \begin{minipage}{0.3\hsize}
    \includegraphics[width=45mm,clip]{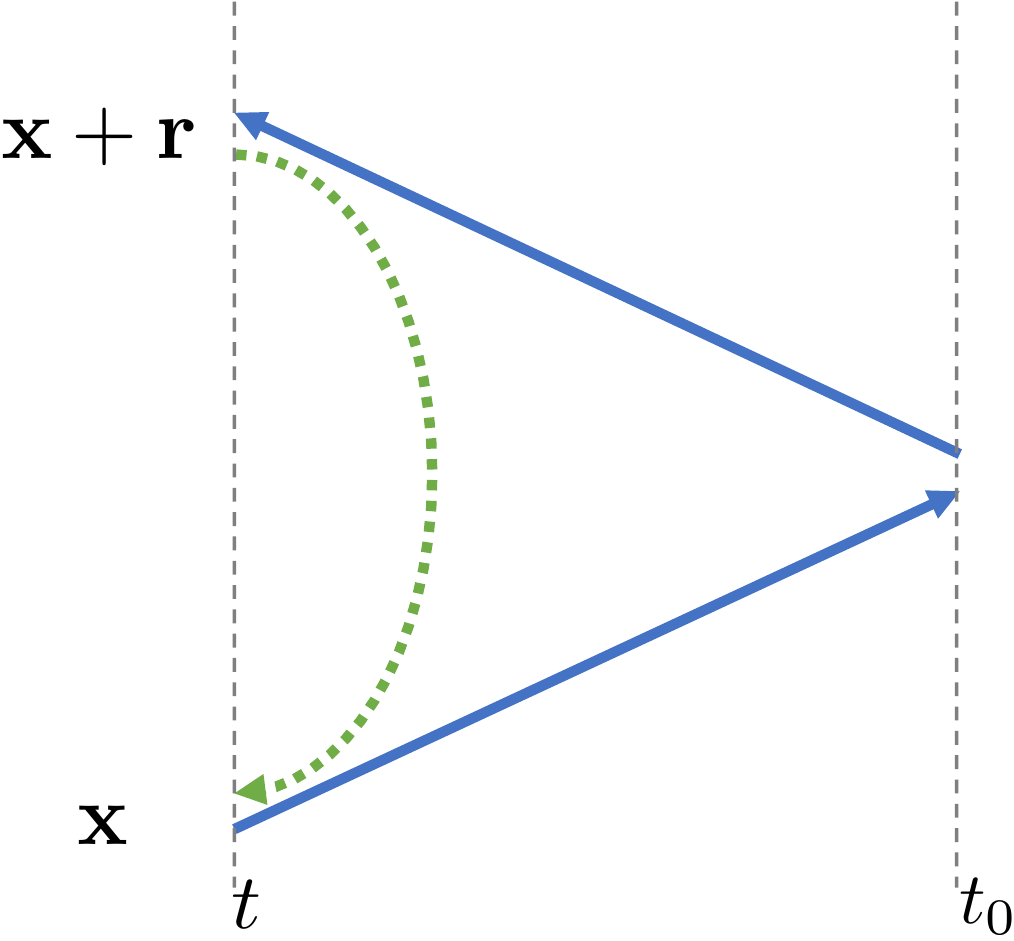}\\
    \centering
    triangle diagram
  \end{minipage}
\end{tabular}
\caption{Representative diagrams appeared in this study.
Blue solid, orange dashed and green dotted lines are calculated with the one-end trick,
sequential propagator and point-to-all propagator, respectively.
Statistical improvement by the CAA is also employed for green dotted lines.
}
\label{fig:diagrams}
\end{figure}

Here we outline details of numerical evaluation for each diagram calculations in this study.
Figure~\ref{fig:diagrams} gives representative quark contraction diagrams appearing in the two-pion correlation functions,
and the techniques utilized in {the evaluations of quark propagator} are shown by different colors and symbols.
Other diagrams similar {to these} representatives are calculated similarly.%in the similar manner.
In the following, we assume to employ a single noise vector for each insertion.
Color and spin indices are implicit for simplicity as well.
%We also employ stochastic estimate for the one-end tricks with one noise vector.
%Color and spin indices  are implicit for simplicity as well.

\subsection{Separated diagram}
As already seen, a separated diagram in Fig.~\ref{fig:diagrams} is written as
\begin{equation}
  \begin{split}
  G^{\rm sep}_{{\bf x};t_0}({\bf r},t) = (+) \sum_{{\bf y_1,y_2}} e^{i{\bf p_z \cdot y_1}} e^{-i{\bf p_z \cdot y_2}} &{\rm tr} \left[  D^{-1} ({\bf x+r},t;{\bf y_1},t_0) \gamma_5 D^{-1} ({\bf y_1},t_0;{\bf x+r},t) \gamma_5 \right] \\
  & \times {\rm tr} \left[ D^{-1} ({\bf x},t;{\bf y_2},t_0) \gamma_5 D^{-1} ({\bf y_2},t_0;{\bf x},t) \gamma_5 \right].
\end{split}
\end{equation}
By using the one-end trick twice, we obtain
\begin{equation}
  G^{\rm sep}_{{\bf x};t_0}({\bf r},t) = (+) \sum_{i,j} \left( \chi_{\gamma_5,t_0[r]}^{(i)\dag}({\bf x+r},t)  \xi_{{\bf p}_z,t_0[r]}^{(i)}({\bf x+r},t) \right) \left( \chi_{\gamma_5,t_0[s]}^{(j)\dag}({\bf x},t)  \xi_{-{\bf p}_z,t_0[s]}^{(j)}({\bf x},t) \right),
\end{equation}
where $i,j$ are indices for dilutions and $r,s$ distinguish independent noise vectors.
In practice, the center-of-mass coordinate ${\bf x}$ is averaged over the whole spacetime to improve the statistical errors,
\begin{equation}
  G^{\rm sep}_{t_0}({\bf r},t) = \frac{1}{L^3} \sum_{\bf x} G^{\rm sep}_{{\bf x};t_0}({\bf r},t).
\end{equation}.

\subsection{Box diagrams}
A box diagram shown in Fig.~\ref{fig:diagrams} is written as
\begin{equation}
  \begin{split}
  G^{\rm box}_{{\bf x};t_0}({\bf r},t) = (-) \sum_{{\bf y_1,y_2}} e^{i{\bf p_z \cdot y_1}} e^{-i{\bf p_z \cdot y_2}} &{\rm tr} [ D^{-1} ({\bf x+r},t;{\bf y_1},t_0) \gamma_5 D^{-1} ({\bf y_1},t_0;{\bf y_2},t_0) \gamma_5 \\
  & \times D^{-1} ({\bf y_2},t_0;{\bf x},t) \gamma_5 D^{-1} ({\bf x},t;{\bf r+x},t) \gamma_5 ]
\end{split}
\end{equation}
For an estimation of this diagram, we first utilize the one-end trick {for} a summation of ${\bf y_2}$,
\begin{equation} \label{eq:boxdiagram_1ststep}
  \begin{split}
  (-) \sum_{i} \sum_{{\bf y_1}} e^{i{\bf p_z \cdot y_1}} &{\rm tr} [ D^{-1} ({\bf x+r},t;{\bf y_1},t_0) \gamma_5
  \xi_{-{\bf p}_z,t_0[r]}^{(i)}({\bf y_1},t_0) \chi_{\gamma_5,t_0[r]}^{(i)\dag}({\bf x},t) D^{-1} ({\bf x},t;{\bf r+x},t) \gamma_5 ].
\end{split}
\end{equation}
We next exactly calculate another all-to-all propagator $D^{-1}({\bf x+r},t;{\bf y_1},t_0)$ by
the sequential propagator technique~\cite{Martinelli:1988rr},
where we consider
a linear equation with a sequential source vector  $e^{i{\bf p_z \cdot y_1}} \gamma_5 \xi_{-{\bf p}_z,t_0[r]}^{(i)}({\bf y_1},t_0)$ as
\begin{equation}
  \left( D \zeta \right) (x) = e^{i{\bf p_z \cdot x}} \gamma_5 \xi_{-{\bf p}_z,t_0[r]}^{(i)}({\bf x},t) \delta_{t,t_0},
\end{equation}
whose solution $\zeta$ is given by
\begin{equation} \label{eq:boxdiagram_sequential}
  \zeta^{(i)}_{{\bf p_z},-{\bf p_z},t_0[r]}({\bf x},t) = \sum_{{\bf y_1}} D^{-1} ({\bf x},t;{\bf y_1},t_0) \gamma_5 \xi_{-{\bf p}_z,t_0[r]}^{(i)}({\bf y_1},t_0) e^{i{\bf p_z \cdot y_1}}.
\end{equation}
Substituting Eq.~(\ref{eq:boxdiagram_sequential}) into Eq.~(\ref{eq:boxdiagram_1ststep}), we obtain
\begin{equation}
  \begin{split}
  G^{\rm box}_{{\bf x};t_0}({\bf r},t) = (-) \sum_{i} \chi_{\gamma_5,t_0[r]}^{(i)\dag}({\bf x},t) H^{-1} ({\bf x},t;{\bf r+x},t) \zeta^{(i)}_{{\bf p_z},-{\bf p_z},t_0[r]}({\bf r+x},t),
\end{split}
\end{equation}
where $H^{-1}$ is an inverse of the hermitized Dirac operator $H = \gamma_5 D$.

To increase statistics of the box diagrams, instead of an average over all ${\bf x}$ with an additional noisy estimation,
we employ the covariant approximation averaging (CAA) for ${\bf x}$,
which is given by
\begin{equation}
  G^{\rm box, imp}_{{\bf x_0};t_0}({\bf r},t) = G^{\rm box, exact}_{{\bf x_0};t_0}({\bf r},t) - G^{\rm box, relaxed}_{{\bf x_0};t_0}({\bf r},t) + \frac{1}{N_G} \sum_{\bf x'} G^{\rm box, relaxed}_{{\bf x'};t_0}({\bf r},t),
\end{equation}
where $N_G$ is the number of a summation over ${\bf x'}$.
Here $G^{\rm box, exact/relaxed}$ is defined as
\begin{equation}
  \begin{split}
    G^{\rm box, exact/relaxed}_{{\bf x_0};t_0}&({\bf r},t)\\
    = (-) \sum_{i} & \Biggl[ \frac{1}{L^3} \sum_{\bf x} \sum_{n}^{N_{\rm low}} \frac{1}{\lambda_n} \chi_{\gamma_5,t_0[r]}^{(i)\dag}({\bf x},t)  v^{(n)}({\bf x},t) v^{(n)\dag}({\bf x+r},t)  \zeta^{(i)}_{{\bf p_z},-{\bf p_z},t_0[r]}({\bf r+x},t) \\
    &+ \chi_{\gamma_5,t_0[r]}^{(i)\dag}({\bf x_0},t) H_{\rm high,exact/relaxed}^{-1} ({\bf x_0},t;{\bf r+x_0},t) \zeta^{(i)}_{{\bf p_z},-{\bf p_z},t_0[r]}({\bf r+x_0},t) \Biggr],
  \end{split}
\end{equation}
where $\lambda_n$ and $v^{(n)}$ are the $n$-th eigenvalue and eigenvector of $H$, respectively, $N_{\rm low}$ is the number of low-eigenmodes used in the CAA, while $H_{\rm high,exact/relaxed}^{-1}$ is an inverse of $H$ projected onto a space
spanned by remaining high-eigenmodes solved with a
tight/relaxed stopping condition.
Since $\chi$ and $\zeta$ are already solved with high precision, we only relax a precision of the sink-to-sink propagator (green dotted line in Fig.~\ref{fig:diagrams}).
Furthermore, we averaged over all ${\bf x}$ in the low-eigenmode part to maximize
statistics.

\subsection{Triangle diagram}
A triangle diagram shown in Fig.~\ref{fig:diagrams} is written as
\begin{equation}
  G^{\rm tri}_{{\bf x};t_0}({\bf r},t) = (-) \sum_{\bf y} {\rm tr} [D^{-1}({\bf r+x},t;{\bf y},t_0) \gamma_3 D^{-1}({\bf y},t_0;{\bf x},t) \gamma_5 D^{-1}({\bf x},t;{\bf r+x},t) \gamma_5].
\end{equation}
Using the one-end trick for a summation over ${\bf y}$, we obtain
\begin{equation}
  \begin{split}
  G^{\rm tri}_{{\bf x};t_0}({\bf r},t) = (-) \sum_{i} \chi_{\gamma_3,t_0[r]}^{(i)\dag}({\bf x},t) H^{-1} ({\bf x},t;{\bf r+x},t) \xi^{(i)}_{{\bf 0},t_0[r]}({\bf r+x},t).
\end{split}
\end{equation}
As in the case of the box diagram, we employ the CAA for ${\bf x}$,
which gives an improved triangle diagram as
\begin{equation}
  G^{\rm tri, imp}_{{\bf x_0};t_0}({\bf r},t) = G^{\rm tri, exact}_{{\bf x_0};t_0}({\bf r},t) - G^{\rm tri, relaxed}_{{\bf x_0};t_0}({\bf r},t) + \frac{1}{N_G} \sum_{\bf x'} G^{\rm tri, relaxed}_{{\bf x'};t_0}({\bf r},t),
\end{equation}
where
\begin{equation}
  \begin{split}
    G^{\rm tri, exact/relaxed}_{{\bf x_0};t_0}&({\bf r},t)\\
    = (-) \sum_{i} & \Biggl[ \frac{1}{L^3} \sum_{\bf x} \sum_{n}^{N_{\rm low}} \frac{1}{\lambda_n} \chi_{\gamma_5,t_0[r]}^{(i)\dag}({\bf x},t)  v^{(n)}({\bf x},t) v^{(n)\dag}({\bf x+r},t)  \xi^{(i)}_{{\bf 0},t_0[r]}({\bf r+x},t) \\
    &+ \chi_{\gamma_5,t_0[r]}^{(i)\dag}({\bf x_0},t) H_{\rm high,exact/relaxed}^{-1} ({\bf x_0},t;{\bf r+x_0},t) \xi^{(i)}_{{\bf 0},t_0[r]}({\bf r+x_0},t) \Biggr].
  \end{split}
\end{equation}

\section{Smeared-sink scheme} \label{appex:smrdsink}
In this appendix, we discuss properties of the smeared-sink scheme in detail.

\subsection{Point-sink scheme vs smeared-sink scheme in $I=1$ $\pi \pi$ system}
\begin{figure}[htbp]
  \hspace{-10mm}
  \begin{tabular}{cc}
  \begin{minipage}{0.5\hsize}
    \includegraphics[width=80mm,clip]{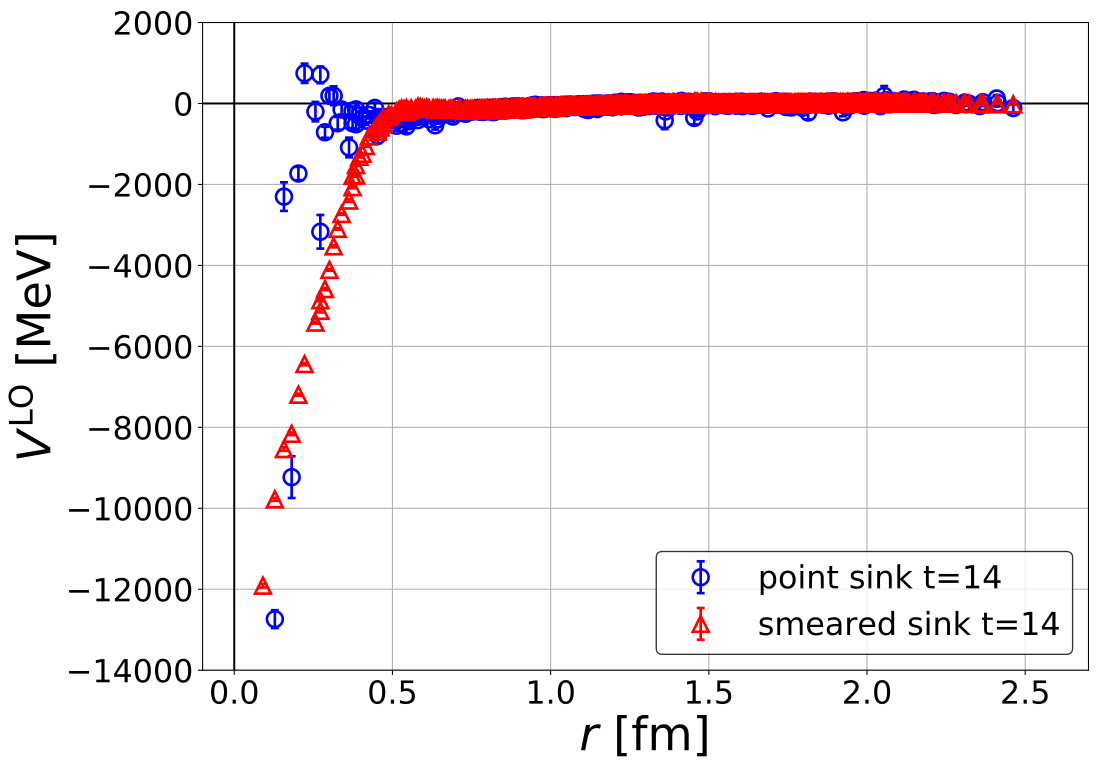}
  \end{minipage} &
  \begin{minipage}{0.5\hsize}
    \includegraphics[width=80mm,clip]{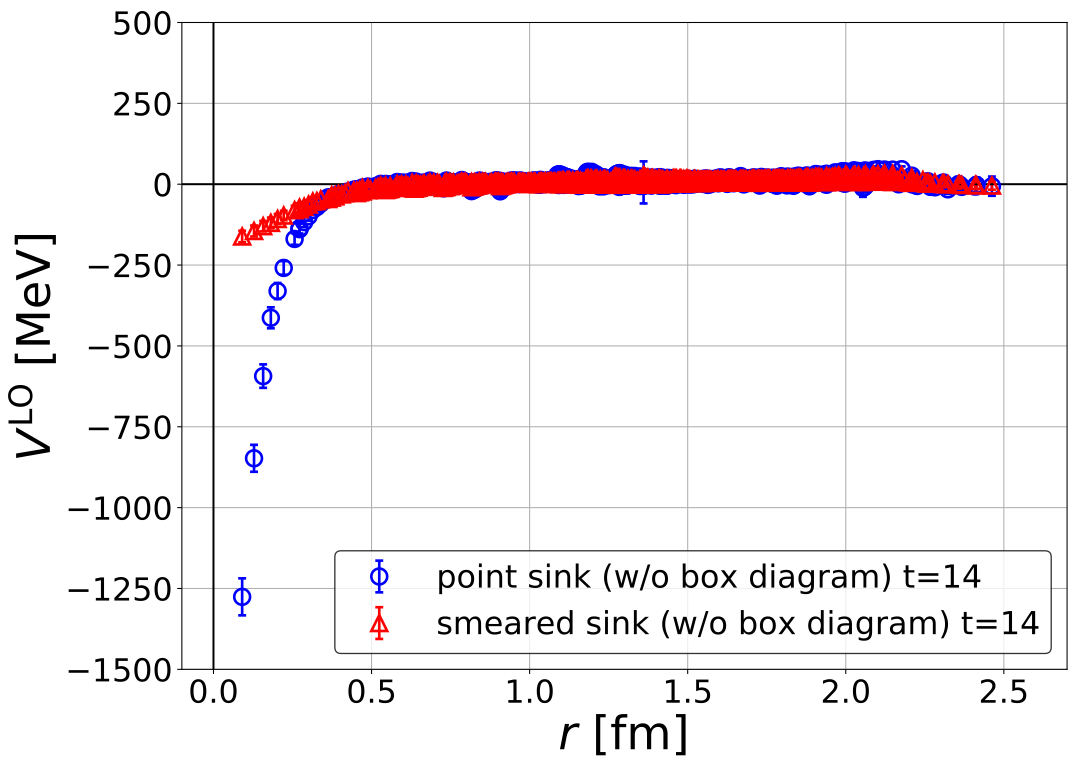}
  \end{minipage}
\end{tabular}
\caption{{A comparison in $I=1$ $\pi\pi$ potential
between two schemes at $t=14$. }
(Left) The effective LO potentials from the $\pi\pi$-type source operator. Blue (red) points show data in the point-sink (smeared-sink) scheme. (Right) Those from the NBS wave function without box diagrams.}
\label{fig:potential_LO_ptvssmr_t14}
\end{figure}

To see why the smeared-sink scheme is needed for the $I=1$ $\pi\pi$ potential, let us compare potentials between the point-sink scheme and the smeared-sink scheme.
Figure~\ref{fig:potential_LO_ptvssmr_t14} (left) shows the $I=1$ $\pi\pi$ potentials calculated from the $\pi \pi$-type source with $N_{\rm conf} = 18$ ($\times$ 64 time slice average).
While the potential in the point-sink scheme
show large non-smooth and scattered behavior at short distances, which makes
a fit to this potential difficult,
such behavior is absent for the potential in the smeared-sink scheme.
Since the potential in the point-sink scheme without box diagrams
does not show such non-smooth behavior(Fig.~\ref{fig:potential_LO_ptvssmr_t14} (right)),
it is probably caused by box diagrams,
which contain quark creation/annihilation processes.

\begin{figure}[htbp]
  \hspace{-10mm}
  \begin{tabular}{cc}
  \begin{minipage}{0.5\hsize}
    \includegraphics[width=80mm,clip]{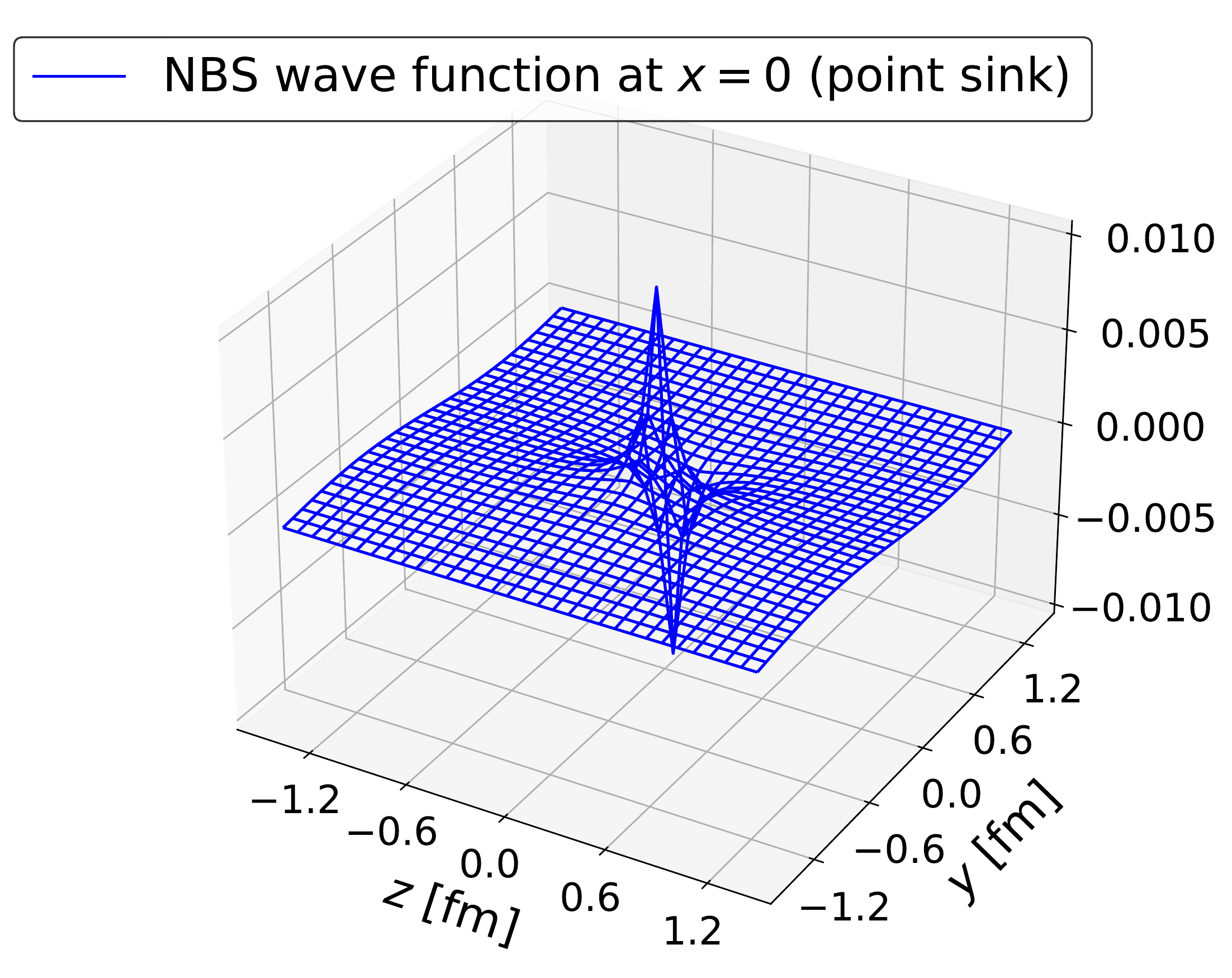}
  \end{minipage} &
  \begin{minipage}{0.5\hsize}
    \includegraphics[width=80mm,clip]{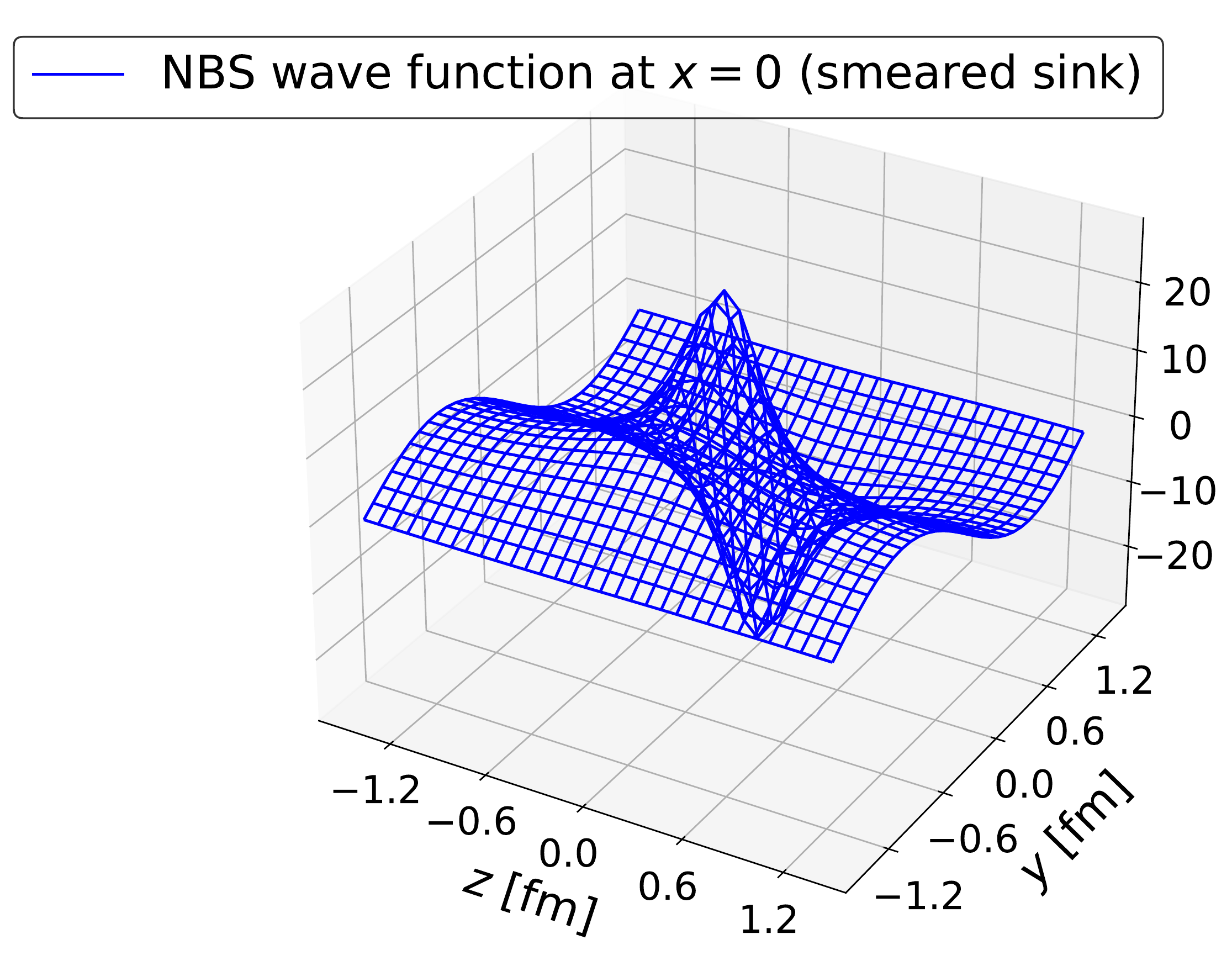}
  \end{minipage}
  \end{tabular}
  \caption{The NBS wave function at $x = 0$ in the point-sink scheme (Left) and the smeared-sink scheme (Right).
  }
  \label{fig:NBS_x0_t14_ptsink}
\end{figure}

We suspect that  this non-smooth and scattered structure is related to a singular behavior of the NBS wave function at short distances, caused by
quark creation/annihilation processes in this channel.
According to the argument by the operator product expansion\cite{Aoki:2010kx,Aoki:2010uz,Aoki:2011aa,Aoki:2012xa,Aoki:2013zj}, the $I=1$ $\pi\pi$ operator at the sink
strongly couples to  the $\rho$ operator at short distance, whose mass dimension is lower than
the $\pi\pi$ operator by 3,
and the NBS wave function behaves as $\psi_{W}({\bf r}) \sim \frac{1}{r^3} Y_{l=1,m=0}(\Omega_{\bf r})$ at short distances.
This implies that the NBS wave function is highly localized and singular around the origin,
which is indeed the case in the point-sink scheme, as seen
in Fig.\ref{fig:NBS_x0_t14_ptsink} (Left).
Since data available around the origin are restricted on a discrete space,
it is difficult to extract a potential smoothly from such a localized wave function
by a discretized Laplacian.
In the smeared-sink scheme, on the other hand,
a singular structure of the {NBS} wave function at short distances is
much milder as seen in Fig.\ref{fig:NBS_x0_t14_ptsink} (Right), so that
the potential reconstructed from discrete data shows a smoother behavior at short distances.
We also expect
similar behaviors of HAL QCD potentials at short distances
generally for other systems which contain quark creation/annihilation diagrams.


\subsection{Effect on the derivative expansion}
The previous HAL QCD study with the LapH method~\cite{Kawai:2017goq} has revealed that the LapH sink-smearing significantly enhances non-localities of HAL QCD
potentials, which makes the derivative expansion less reliable.
Therefore we would like to check whether our sink-smearing scheme given in eq.(\ref{eq:sinkop}) is free from such a problem.
For this purpose,
we calculate $I=2$ $\pi \pi$ potential in both point-sink and smeared-sink schemes and compare LO phase shifts between the two schemes.

Calculations of NBS wave functions in both schemes are performed by using the one-end trick with full color/spin dilution and $s2$ space dilution for a single $Z_4$ noise.
A number of configuration is $N_{\rm conf} = 10$ ($\times 64$ timeslice average), and statistical errors are estimated by the jackknife method with bin-size 1.

\begin{figure}[htbp]
  \hspace{-10mm}
  \begin{tabular}{cc}
  \begin{minipage}{0.5\hsize}
    \includegraphics[width=80mm,clip]{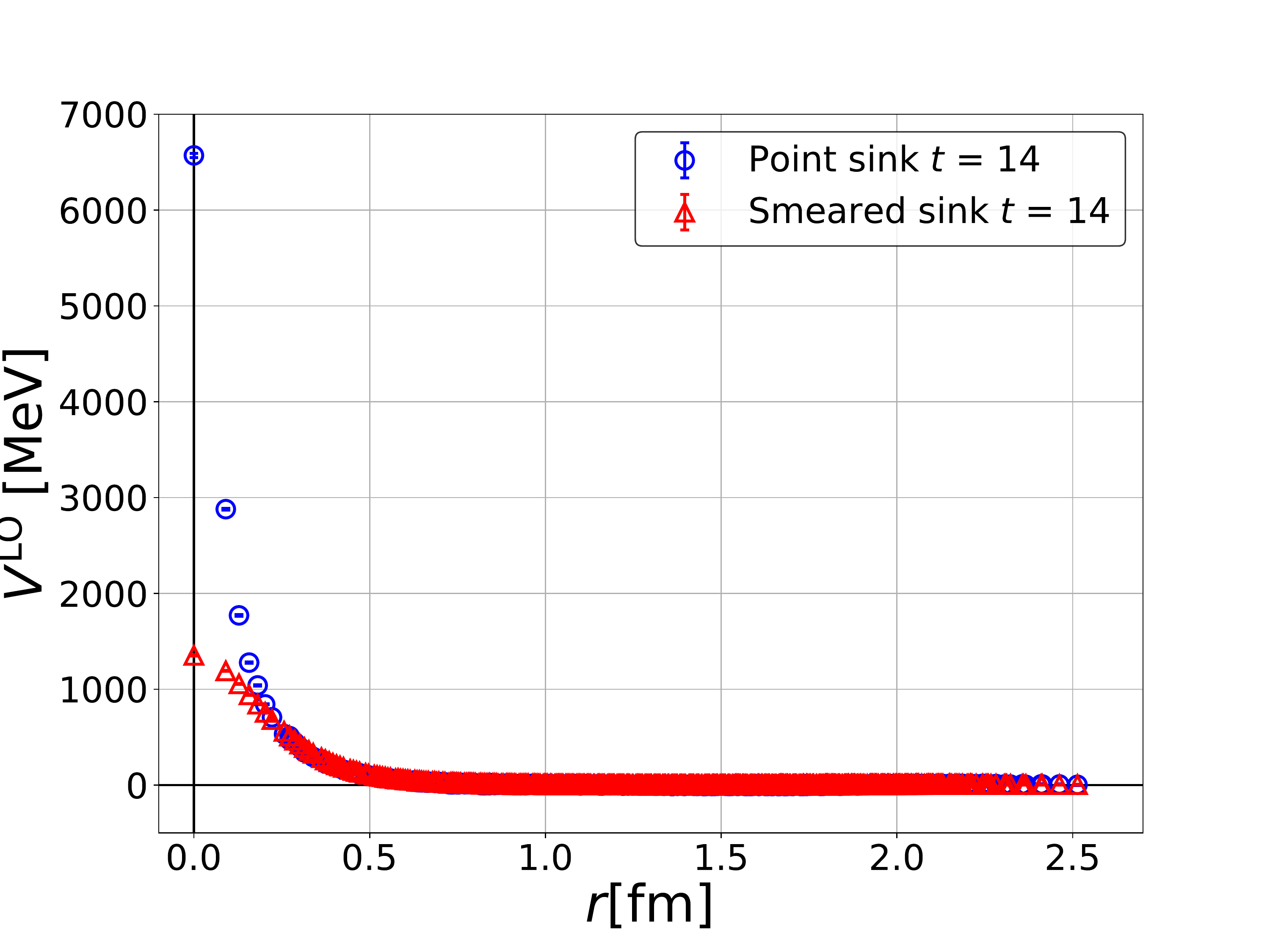}
  \end{minipage} &
  \begin{minipage}{0.5\hsize}
    \includegraphics[width=80mm,clip]{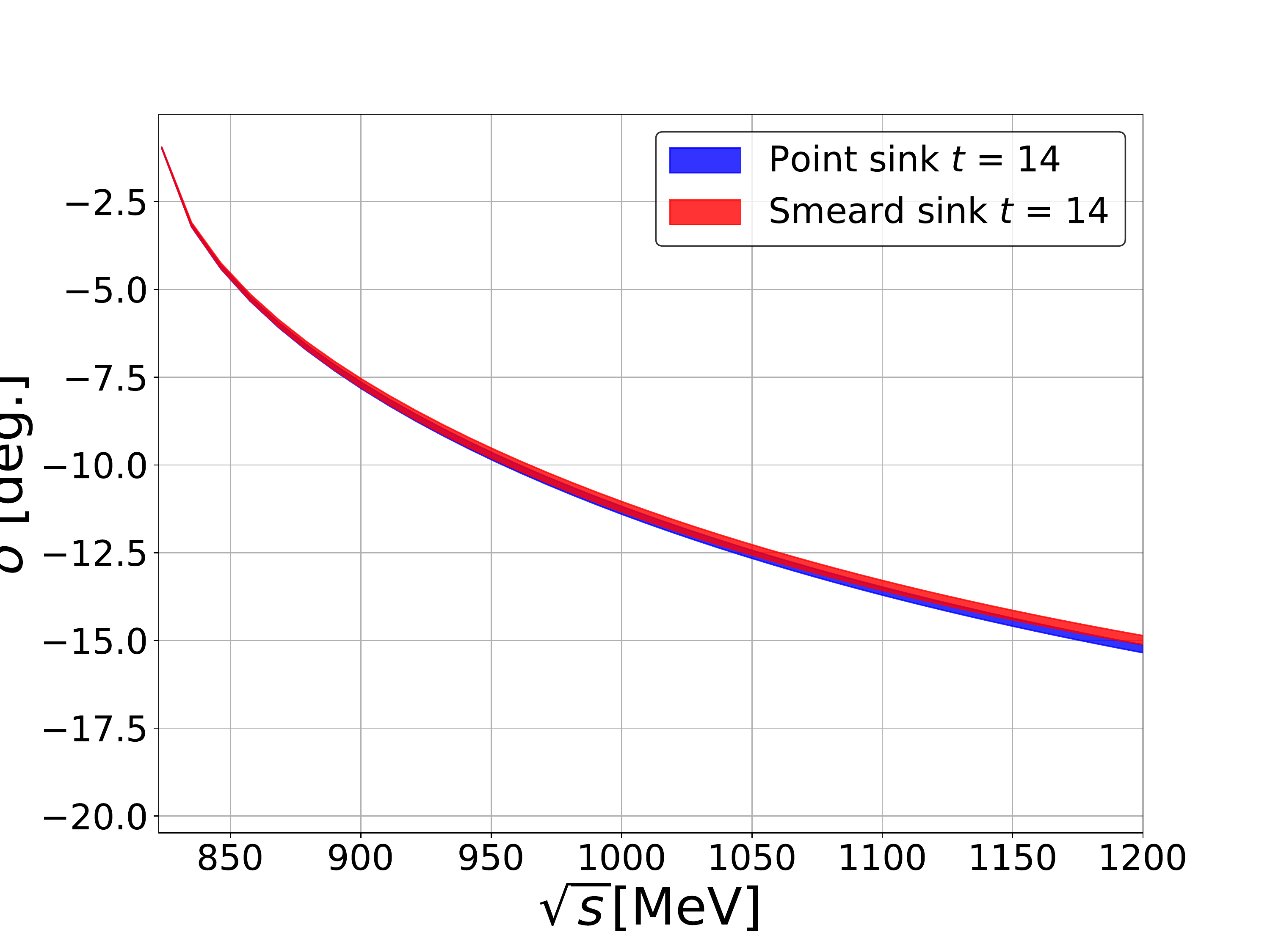}
  \end{minipage}
\end{tabular}
\caption{{A comparison between point-sink and smeared-sink schemes for}
the $I=2$ $\pi\pi$ system. (Left) Effective LO potentials. Blue (red) points show data in the point-sink (smeared-sink) scheme. (Right) {Corresponding} phase shifts.}
\label{fig:i2ppLO_smrdsink}
\end{figure}

Figure~\ref{fig:i2ppLO_smrdsink}(left) shows effective LO potentials at $t = 14$.
Potentials between two schemes show significantly different behaviors only at short distances,
which however do not affect  phase shifts at $\sqrt{s} < 1200$ MeV,
as plotted in Fig.~\ref{fig:i2ppLO_smrdsink} (right).
Thus the smeared-sink
  does not enhance non-locality of the $I=2$ $\pi\pi$ potential in this energy region.
Since a relevant energy range for the $\rho$ resonance in this study is well covered
by this energy region ($\sqrt{s} < 1200$ MeV),
we also expect that non-locality of the $I=1$ $\pi\pi$ potential is not enhanced by the smeared-sink scheme, either.

\section{Assumptions for the analysis of $V_2^{\rm N^2LO}$ } \label{appex:fitassumption}
In this appendix, we discuss assumptions made for the analysis of $V_2^{\rm N^2LO}$.
The effective LO potential is related to the exact non-local potential as
\begin{equation}
  V_i^{\rm LO} = V_0 + V_2 \frac{\nabla^2 R_i}{R_i} + V_4 \frac{\nabla^4 R_i}{R_i} + \cdots.
\end{equation}
Let us consider a case where the N$^4$LO and higher terms are negligibly small. In this case, the effective LO potential and the exact N$^2$LO potential $V_0, V_2$ can be related by
\begin{equation}
  V_i^{\rm LO} = V_0 + V_2 \frac{\nabla^2 R_i}{R_i},\quad {(i=A,B)},
\end{equation}
which leads to
\begin{equation}
  \Delta V^{\rm LO} \equiv V_A^{\rm LO} - V_B^{\rm LO} = V_2 \left( \frac{\nabla^2 R_A}{R_A} - \frac{\nabla^2 R_B}{R_B} \right) \equiv V_2 \Delta \left( \frac{\nabla^2 R}{R} \right).
  \label{eq:V2NLO}
\end{equation}
Using this relation, we find
\begin{equation}
  \Delta \left( \frac{\nabla^2 R}{R} \right) = 0 \ {\rm at}\ r = r_0 \ \Rightarrow\  \Delta V^{\rm LO} = 0\ {\rm and}\ \Delta V_{\rm ene}^{\rm LO} = 0 \ {\rm at}\ r = r_0,
\end{equation}
where $\Delta V_{\rm ene}^{\rm LO} \equiv \Delta V^{\rm LO} - \Delta \left( \frac{\nabla^2 R}{R} \right) / 2 \mu$.
Therefore, if $\Delta \left( \frac{\nabla^2 R}{R} \right)$ vanishes
at $r_0, r_1, \cdots$, both $\Delta V^{\rm LO}$ and $\Delta V_{\rm ene}^{\rm LO}$ must become zero also at those points.

\begin{figure}[htbp]
  \centering
  \includegraphics[width=80mm,clip]{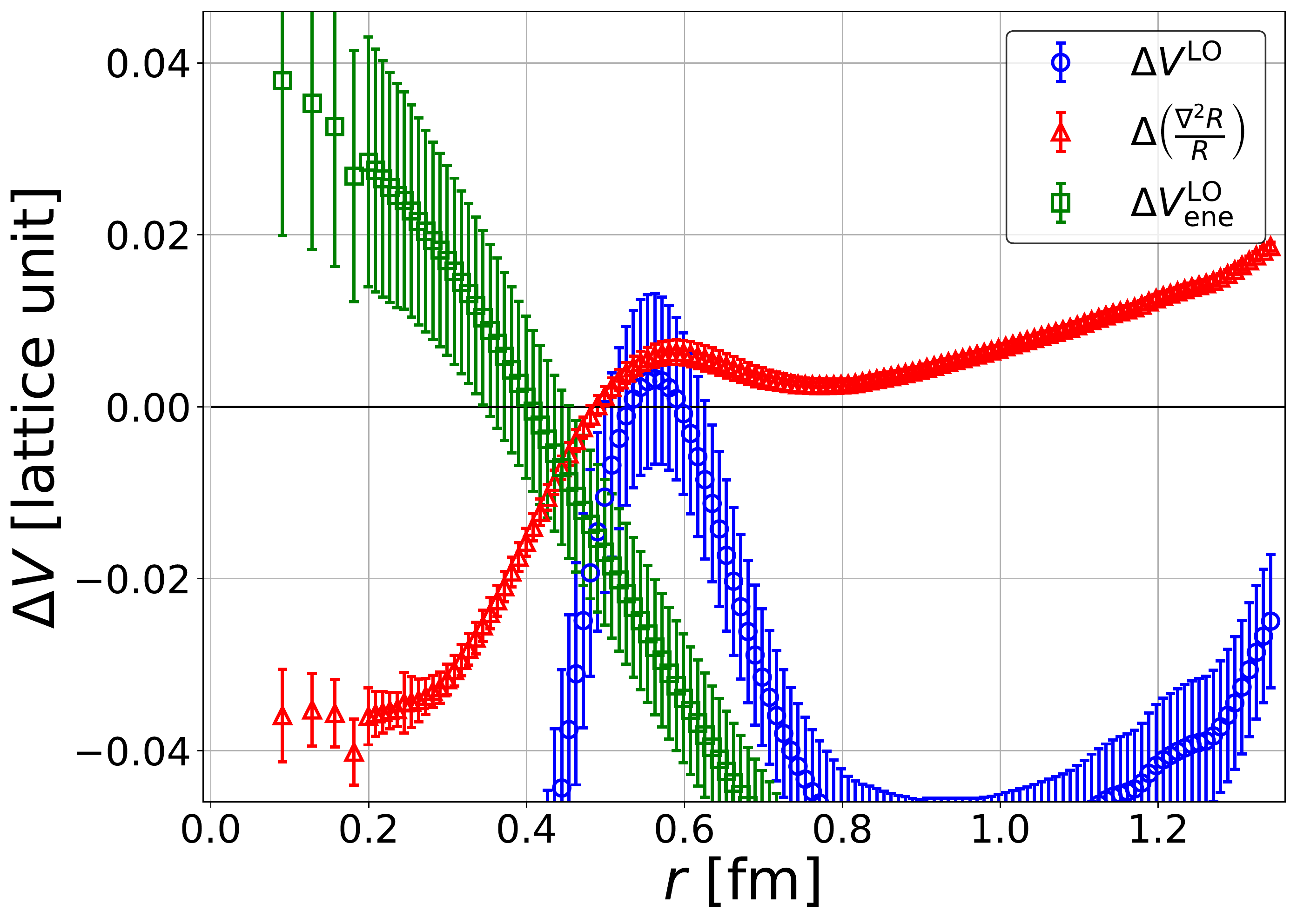}
  \caption{Behaviors of $\Delta V^{\rm LO}$(blue circles), $\Delta (\nabla^2 R / R)$(red triangles) and $\Delta V_{\rm ene}^{\rm LO}$(green squares).}
  \label{fig:deltaLOpots}
\end{figure}

Figure~\ref{fig:deltaLOpots} shows data of $\Delta V^{\rm LO}$, $\Delta \left( \frac{\nabla^2 R}{R} \right)$, and $\Delta V_{\rm ene}^{\rm LO}$ in this study.
While $\Delta \left( \frac{\nabla^2 R}{R} \right)$ has a single zero, $\Delta V^{\rm LO}$ and
$\Delta V_{\rm ene}^{\rm LO}$ have zeros at slightly different positions,
probably due to the neglected higher order effects in $\Delta V^{\rm LO}$ and $\Delta V_{\rm ene}^{\rm LO}$.
We assume in our N$^2$LO analysis that our data are well described
without N$^4$LO and {higher order} terms
so that
$\Delta V^{\rm LO}$, $\Delta \left( \frac{\nabla^2 R}{R} \right)$ and $\Delta V_{\rm ene}^{\rm LO}$ share a common zero point.
This assumption motivates us to employ a non-singular function which satisfies
$2\mu V_2^{\rm N^2LO} - 1\ (= 2\mu \Delta V_{\rm ene}^{\rm LO}/\Delta\left(\frac{\nabla^2 R}{R}\right))\ < 0$
at all $r$ in the fit of $V_2^{\rm N^2LO}$.

\section{Energy-dependent local N$^2$LO potential} \label{appex:enedeplocalpot}
\begin{figure}[tbp]
  \hspace{-10mm}
  \begin{tabular}{cc}
  \begin{minipage}{0.5\hsize}
    \includegraphics[width=80mm,clip]{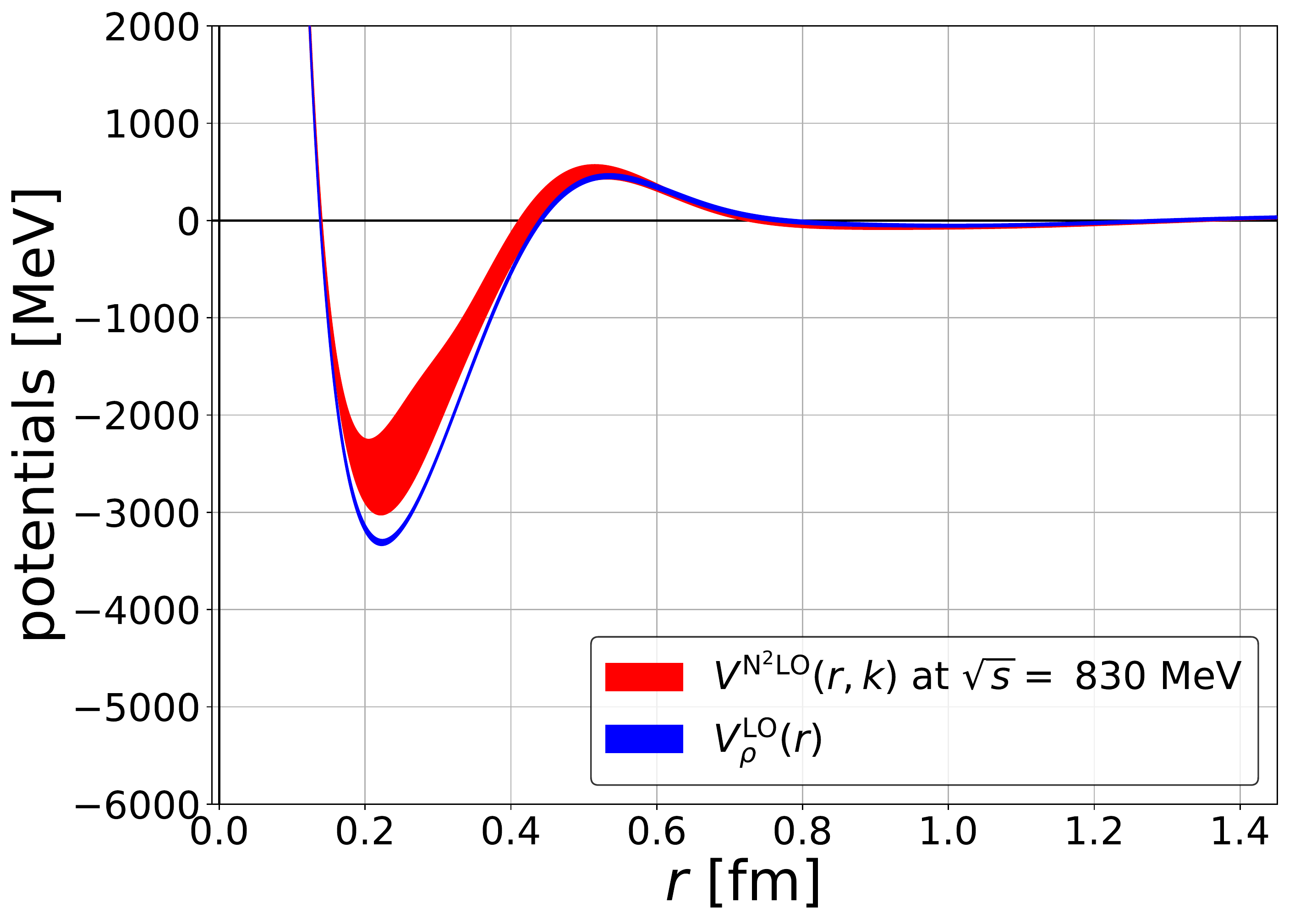}
  \end{minipage} &
  \begin{minipage}{0.5\hsize}
    \includegraphics[width=80mm,clip]{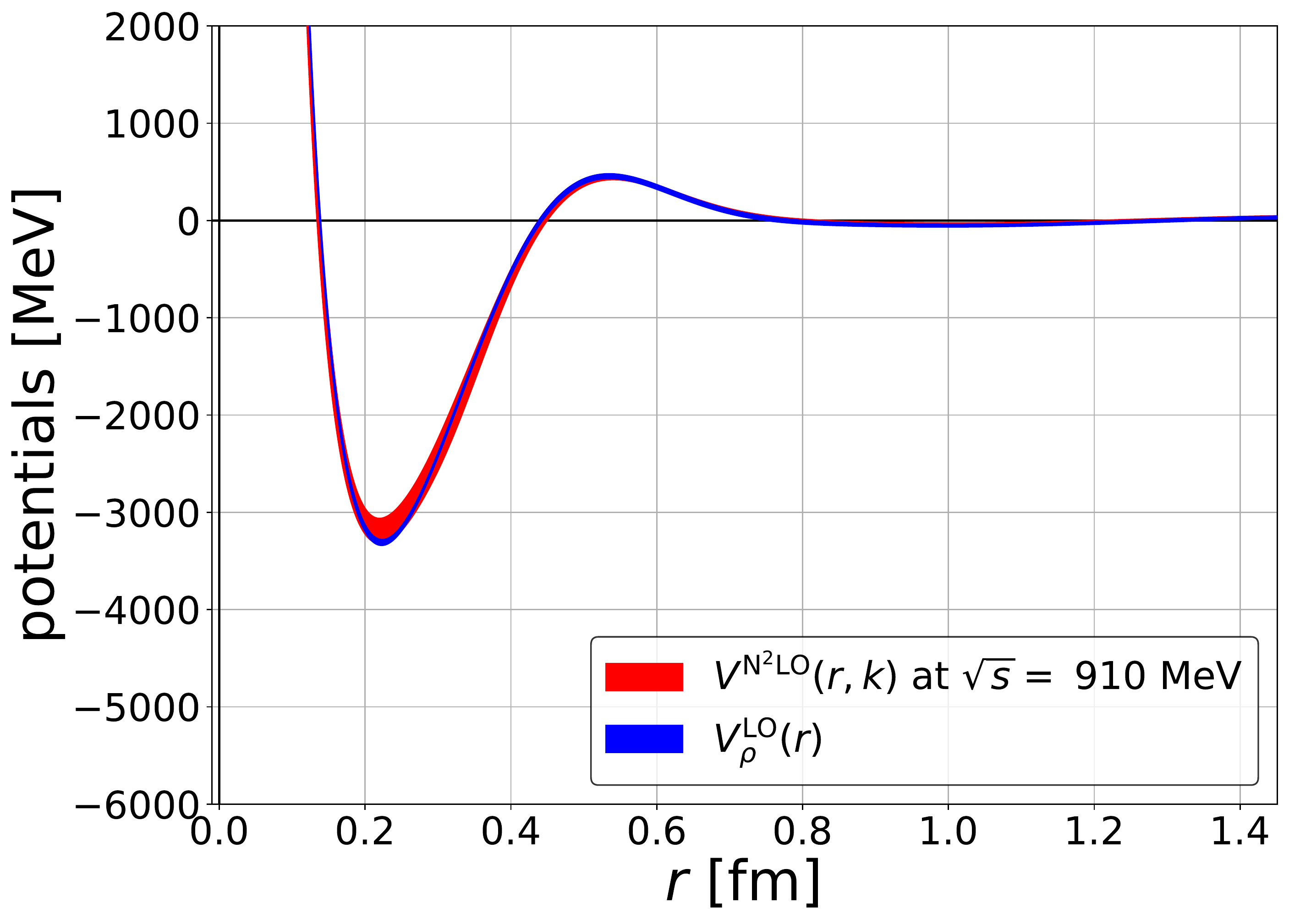}
  \end{minipage} \\
  \begin{minipage}{0.5\hsize}
    \includegraphics[width=80mm,clip]{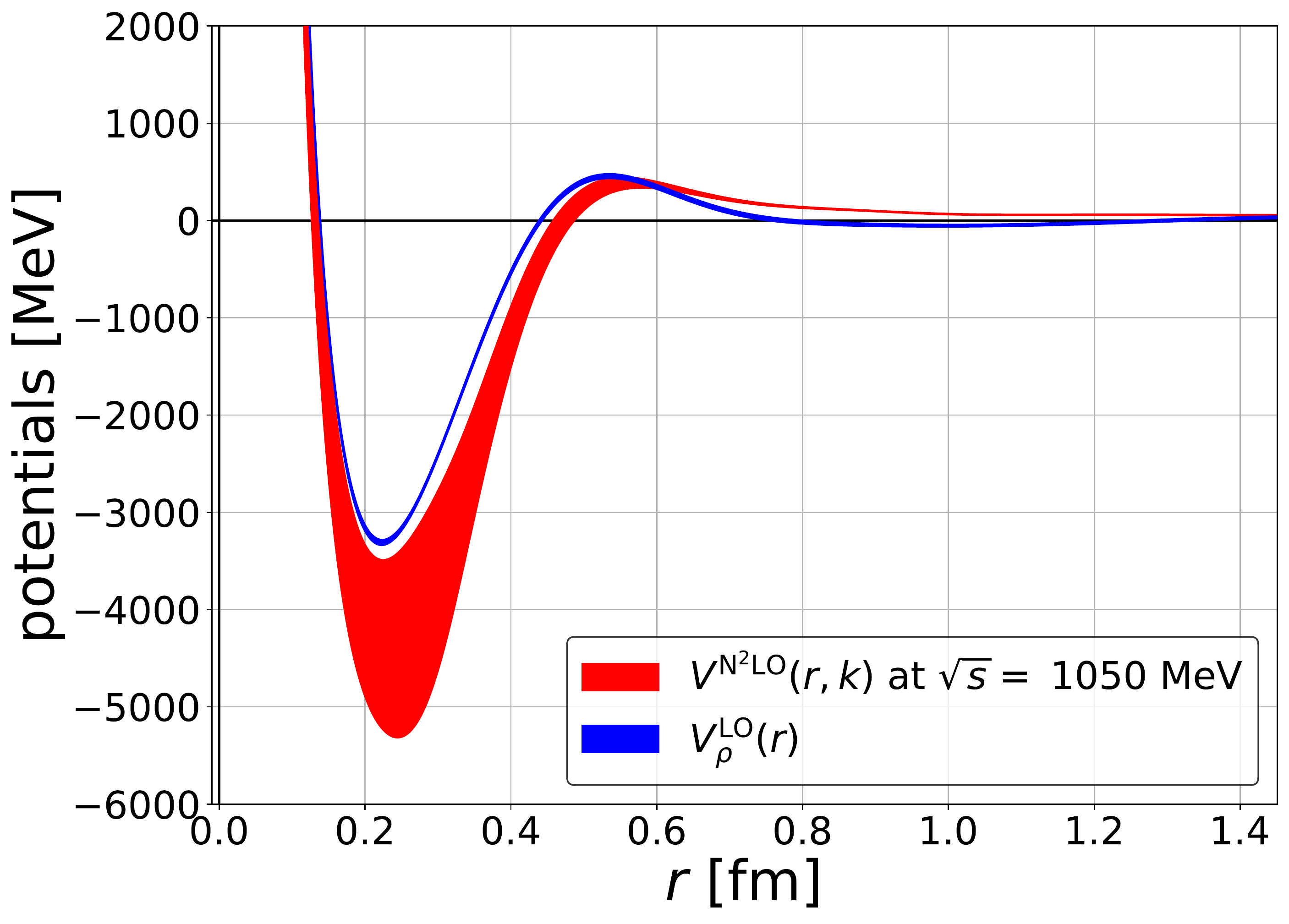}
  \end{minipage}
  \end{tabular}
  \caption{Energy-dependent local N$^2$LO potentials. (Upper left) near threshold ($\sqrt{s} = 830$ MeV). (Upper right) near the CM frame ground state energy ($\sqrt{s} = 910$ MeV). (Lower left) larger energy ($\sqrt{s} = 1050$ MeV). For a comparison, we show the effective LO potential with $\rho$-type source. }
  \label{fig:enedep_potentials}
\end{figure}

Here, we discuss our N$^2$LO potential in a different point of view, an energy-dependent local form.
We can convert the energy-independent non-local N$^2$LO potential $U^{\rm N^2LO} = V^{\rm N^2LO}_0 + V^{\rm N^2LO}_2 \nabla^2$ to an energy-dependent local form $V^{\rm N^2LO}(r;k)$ as~\cite{Iritani:2018zbt}
\begin{equation}
  V^{\rm N^2LO}(r;k) = \frac{V^{\rm N^2LO}_0 - k^2 V^{\rm N^2LO}_2}{1 - m_{\pi} V^{\rm N^2LO}_2}.
\end{equation}
Figure~\ref{fig:enedep_potentials} shows this energy-dependent local potentials
with the centrifugal term
at several energies: near threshold ($\sqrt{s} = 830$ MeV), near the ground state energy in the center-of-mass frame ($\sqrt{s} = 910$ MeV),
and at higher energy ($\sqrt{s} = 1050$ MeV).
At low energies,
we observe that the attractive pocket of the $V^{\rm N^2LO}(r;k)$ is smaller than
that of the LO potential $V^{\rm LO}_{\rho}$ which
makes N$^2$LO phase shifts smaller than LO phase shifts.
Around the CM ground state energy,
$V^{\rm N^2LO}(r;k)$ and $V^{\rm LO}_{\rho}$
are almost identical, since $V^{\rm LO}_{\rho}$ is obtained from correlators saturated by that state.
At high-energy region, a difference between $V^{\rm N^2LO}(r;k)$ and $V^{\rm LO}_{\rho}$ becomes larger in all ranges.
The significant improvement by the N$^2$LO analysis for the phase shifts  at high energies
can be understood from this difference.

%%% references (using BibTeX) %%%
%\bibliographystyle{unsrt}
\bibliography{ref}
\bibliographystyle{apsrev4-1.bst}

\end{document}